\documentclass[fleqn,usenatbib]{mnras}

\usepackage{newtxtext,newtxmath}

\usepackage{natbib}
\usepackage{multicol}

\usepackage[T1]{fontenc}
\usepackage{ae,aecompl}

\usepackage{graphicx}	
\usepackage{amsmath}	
\usepackage{amssymb}	
\usepackage{enumerate}

\newcommand{\uvec}[1]{\boldsymbol{\hat{\textbf{#1}}}}
\newcommand{\dlim}{d_\mathrm{host}\leq300\mathrm{\,kpc}}
\newcommand{\dlimand}{d_\mathrm{host, proj}\leq150\mathrm{\,kpc}}
\newcommand{\mstarlowlim}{M_*\geq10^5\mathrm{\,\Msun}}
\newcommand{\mpeaklowlim}{M_\mathrm{peak}\geq8\times10^8\mathrm{\,\Msun}}
\newcommand{\rninetyfifty}{R_{90}/R_{50}}
\newcommand{\rninetyten}{R_{90}/R_{10}}
\newcommand{\rmsh}{\Delta_\mathrm{h}}

\newcommand{\orb}{\Delta_\mathrm{orb}}
\newcommand{\zzero}{z=0}
\newcommand{\zzerotwo}{z=0-0.2}
\newcommand{\zzerofive}{z=0-0.5}

\newcommand{\nitertenk}{10^4}

\newcommand{\vlos}{v_\mathrm{los}}
\newcommand{\nsat}{N_\mathrm{sat}}
\newcommand{\nsatft}{N_\mathrm{sat}=14}
\newcommand{\mystar}{*}
\newcommand{\Msun}{\mathrm{M}_{\odot}}
\newcommand{\Mstar}{M_{\mystar}}

\newcommand{\Mhalo}{M_\mathrm{halo}}
\newcommand{\Mtwohm}{M_\mathrm{200m}}
\newcommand{\Mpeak}{M_\mathrm{peak}}
\newcommand{\lcdm}{\Lambda\mathrm{CDM}}
\newcommand{\Gaia}{\textit{Gaia}}
\newcommand{\vpeak}{V_\mathrm{peak}}
\newcommand{\vcirc}{V_\mathrm{circ}}
\newcommand{\fmax}{f^{\rm max}_{\vlos}}

\newcommand{\degree}{^\circ}

\title[Satellite planes in FIRE]{Planes of satellites around Milky Way/M31-mass galaxies in the FIRE simulations and comparisons with the Local Group}

\author[J. Samuel et al.]{
Jenna Samuel$^{1}$\thanks{E-mail: jsamuel@ucdavis.edu},
Andrew Wetzel$^{1}$,
Sierra Chapman$^{1}$,
Erik Tollerud$^{2}$,
\newauthor
Philip F. Hopkins$^{3}$,
Michael Boylan-Kolchin$^{4}$,
Jeremy Bailin$^{5}$,
\newauthor
Claude-Andr{\'e} Faucher-Gigu{\`e}re$^{6}$
\\
$^{1}$Department of Physics and Astronomy, University of California, Davis, CA 95616, USA\\
$^{2}$Space Telescope Science Institute, 3700 San Martin Dr, Baltimore, MD 21218, USA\\
$^{3}${TAPIR, Mailcode 350-17, California Institute of Technology, Pasadena, CA 91125, USA}\\
$^{4}${Department of Astronomy, The University of Texas at Austin, 2515 Speedway, Stop C1400, Austin, TX 78712, USA}\\
$^{5}${Department of Physics and Astronomy, University of Alabama, Box 870324, Tuscaloosa, AL 35487-0324, USA}\\
$^{6}${Department of Physics and Astronomy and CIERA, Northwestern University, 2145 Sheridan Road, Evanston, IL 60208, USA}\\
}

\pubyear{2020}

\begin{document}
\label{firstpage}
\pagerange{\pageref{firstpage}--\pageref{lastpage}}
\maketitle

\begin{abstract}

We examine the prevalence, longevity, and causes of planes of satellite dwarf galaxies, as observed in the Local Group.
We use 14 Milky Way/Andromeda-(MW/M31) mass host galaxies from the FIRE-2 simulations.
We select the 14 most massive satellites by stellar mass within $\dlim$ of each host and correct for incompleteness from the foreground galactic disc when comparing to the MW.
We find that MW-like planes as spatially thin and/or kinematically coherent as observed are uncommon, but they do exist in our simulations.
Spatially thin planes occur in 1--2 per cent of snapshots during $\zzerotwo$, and kinematically coherent planes occur in 5 per cent of snapshots.
These planes are generally transient, surviving for $<500$ Myr.
However, if we select hosts with an LMC-like satellite near first pericenter, the fraction of snapshots with MW-like planes increases dramatically to $7-16$ per cent, with lifetimes of $0.7-3$ Gyr, likely because of group accretion of satellites.
We find that M31's satellite distribution is much more common: M31's satellites lie within $\sim1\sigma$ of the simulation median for every plane metric we consider.
We find no significant difference in average satellite planarity for isolated hosts versus hosts in LG-like pairs.
Baryonic and dark matter-only simulations exhibit similar levels of planarity, even though baryonic subhaloes are less centrally concentrated within their host halos.
We conclude that planes of satellites are not a strong challenge to $\lcdm$ cosmology.

\end{abstract}




\section{Introduction}\label{intro}

Astrometric measurements have revealed that a subset of the Milky Way (MW) satellite galaxies coherently orbit their host galaxy within a spatially thin plane (`thin' describes systems with minor-to-major axis ratios of $c/a \lesssim 0.3$, and `coherent' indicates that a majority of satellites share the same orbital direction) \citep[e.g.,][]{LyndenBell1976,Kroupa2005,Pawlowski2012a}.
Recently, precise proper motions from $\Gaia$ Data Release 2 have affirmed an even tighter orbital alignment of MW satellites than previously measured \citep{Fritz2018,Pawlowski2020}.
Similar structures have also been observed around Andromeda (M31) \citep{Ibata2013,Conn2013} and Centaurus A \citep{Muller2018}. 
However, the spatial and kinematic coherence of satellite planes beyond the Local Group (LG) is less certain because of projection effects, distance uncertainties, and the inaccessibility of proper motions.
Even at the relatively close distance of M31, currently only two of its satellites have measured proper motions \citep{Sohn2020}, making it difficult to determine true 3D orbital alignment of the entire satellite population.

The cosmological significance of these satellite planes remains a topic of ongoing investigation, largely because of a lack of consensus on the incidence of planarity in both simulations and observations.
Studies using dark matter-only (DMO) simulations have often yielded conflicting interpretations of how rare satellite planes are in the standard cosmological model of cold dark matter with a cosmological constant ($\lcdm$).
Most analyses of DMO simulations find such configurations to be rare, highly significant, and therefore possibly in conflict with $\lcdm$ \citep[e.g.][]{Metz2008,Pawlowski2014,Buck2016}.
However, DMO simulations combined with semi-analytic models of galaxy formation suggest that planes might be more common \citep{Libeskind2009,Cautun2015}, but this is not a universal result \citep{Pawlowski2014b,Ibata2014a}.
Results from baryonic simulations have varied too, often relying on a much smaller sample of host-satellite systems compared to what is available from DMO simulations.
Some baryonic simulations show evidence for a more natural presence of satellite planes in the universe \citep[e.g.][]{Libeskind2007,Sawala2016}.
While other baryonic results show that satellite planes can be uncommon, but find conflicting evidence for whether planes can be explained by anisotropic satellite accretion along filamentary structures \citep{Ahmed2017,Shao2018,Shao2019}.

Beyond just checking for the presence and significance of satellite planes in simulations, several authors have also explored what may cause planes to form, with mixed results.
Though one might expect the host halo to affect satellite planes,
\citet{Pawlowski2014} found no connection between planes and host halo properties.
Some authors have argued either for \citep{Zentner2005,Libeskind2011} or against \citep{Pawlowski2012b} the preferential infall of satellites along cosmic filaments as a causal factor in the formation of satellite planes.
\citet{Li2008} proposed the accretion of satellites in small groups as an explanation of correlated orbits, and \citet{Wetzel2015a} showed that $25-50$ per cent of satellite dwarf galaxies in MW-mass hosts today previously were part of a group.
\citet{Metz2007} even speculated that satellite planes arise naturally from the creation of tidal dwarf galaxies in fly-bys or mergers of larger galaxies.

Several authors have investigated the orbital stability of LG satellite planes.
Recently, \citet{Riley2020} showed that globular clusters and stellar streams around the MW do not seem to be members of the satellite plane, suggesting that plane members may be recently accreted or in a particularly stable orbital configuration.
\citet{Pawlowski2017} noted that integrating present-day satellite orbits either forward or backward in time typically leads to the disintegration of the plane, especially when sampling measurement uncertainties on satellite galaxy positions and velocities.
\citet{Shaya2013} took a different approach and, by searching the dynamical parameter space of Local Volume satellites, found past trajectories that could possibly lead to the observed satellite planes.

Many previous attempts to investigate satellite planes have relied on simulations that may not resolve the dynamical evolution of ``classical'' ($\mstarlowlim$) dwarf galaxies, or that do not include baryonic physics.
Insufficient resolution can lead to artificial satellite destruction \citep[e.g.][]{Carlberg1994,vanKampen1995,Moore1996,Klypin1999a,vanKampen2000,Diemand2007,Wetzel2010,vandenBosch2018}.
This may introduce a bias in satellite plane metrics if the destruction is spatially varying (such as near the host disc), and because earlier infalling satellites are preferentially destroyed, leading to an age bias that correlates with satellite orbit today \citep{Wetzel2015a}.

If baryonic effects act to create or destroy planes of satellites, then dark matter-only simulations may not be able to wholly capture the theoretical picture of satellite plane formation.
The central disc in baryonic simulations tidally destroys satellites, altering their radial profile at small distances from the host \citep[e.g.,][]{DOnghia2010,Sawala2017,GK2017b,Nadler2018,Kelley2018,RodriguezWimberly2019,Samuel2020}.
This leads the surviving satellites to have more tangentially biased orbits \citep{GK2017b,GK2019a}, but these effects do not necessarily imply an effect on planarity.
In addition, \citet{Ahmed2017} found that the members of satellite planes in baryonic versus DMO simulations of the same host halo can be different, suggesting that baryonic effects may alter halo occupation in unexpected ways and hence affect satellite planes.
\citet{GK2019a} also noted that satellites in baryonic simulations of LG-like pairs do not necessarily trace the most massive subhaloes in DMO runs of the same systems.

Outside of the MW, the satellite plane around M31 is somewhat more ambiguous.
Taken as a whole, M31's satellites do not appear to be particularly planar, but a subset of 15 satellites lie within a significantly spatially thin plane and most of those are kinematically aligned, based on line-of-sight velocities \citep{Conn2013,Ibata2013}.
Many works have focused in on this particular subset, but it is important to understand the overall satellite distribution, because there are no clear evolutionary differences between M31 plane members and non-members \citep{Collins2015}.

Satellite planes outside of the LG are more difficult to robustly characterize because of projection effects and larger distance uncertainties.
Studies using the Sloan Digital Sky Survey (SDSS) database have revealed that while there is evidence for spatial flattening of satellites \citep[e.g.,][]{Brainerd2005}, their kinematic distribution is unlikely to indicate a coherently orbiting satellite plane \citep{Phillips2015}.
Furthermore, the Satellites Around Galactic analogues (SAGA) survey \citep{Geha2017}, which aims to study satellites of $\sim100$ MW analogues in the nearby Universe, has found little evidence for coherently orbiting satellite planes \citep{Mao2020}.

In this paper, we seek to understand if the FIRE-2 simulations contain satellite planes similar to those found in the Local Group, whether those satellite planes are long-lived or transient, and if the presence of satellite planes correlates with host or satellite properties.
We leave comparisons to systems outside of the LG for future work.
We organize this paper as follows: in Section~\ref{sims} we describe our simulations and satellite selection criteria, in Section~\ref{observations} we describe the 3D positions and velocities of Local Group satellites used, in Section~\ref{methods} we describe the plane metrics we apply to simulations and observations, in Section~\ref{results} we present our results of planarity in simulations compared to observations, and in Section~\ref{discussion} we discuss our conclusions and their implications for observed satellite planes.

\section{Simulations}
\label{sims}

The zoom-in simulations we use in this work reproduce the mass functions, radial distributions, and star formation histories of classical ($\mstarlowlim$) dwarf galaxies around MW/M31-like hosts \citep{Wetzel2016,GK2019a,GK2019b,Samuel2020}.

We use two suites of cosmological zoom-in hydrodynamic simulations from the Feedback In Realistic Environments (FIRE) project\footnote{\url{https://fire.northwestern.edu/}}.
Latte is currently a suite of 7 isolated MW/M31-mass galaxies with halo masses M$_{\rm 200m} = 1 - 2 \times 10^{12}\, \Msun$\footnote{`200m' indicates a measurement relative to 200 times the mean matter density of the Universe} introduced in \citet{Wetzel2016}.
We selected the Latte halos for zoom-in re-simulation from a periodic volume dark matter simulation box of side length 85.5 Mpc.
We selected two of the Latte halos (m12r and m12w) to host an LMC-mass subhalo at $\zzero$ within their initial DMO simulations, though after re-simulation with baryonic physics the orbital phase of these subhaloes changes and they are no longer near pericenter \citep{Samuel2020}.
Latte gas and star particles have initial masses of $7070\, \Msun$, but at $z = 0$ a typical star particle has mass $\approx 5000\, \Msun$ because of stellar mass loss.
Dark matter particles have a mass resolution of m$_{\rm dm} = 3.5 \times 10^4\, \Msun$.
The gravitational softenings (comoving at $z > 9$ and physical at $z < 9$) of dark matter and stars particles are fixed: $\epsilon_{\rm dm} = 40$ pc and $\epsilon_{\rm star} = 4$ pc (Plummer equivalent). 
The gas softening is fully adaptive, matched to the hydrodynamic resolution, and the minimum gas resolution (inter-element spacing) and softening length reached in Latte is $\approx 1$ pc.
We also use an additional simulation of an isolated MW/M31-mass galaxy (m12z), simulated at higher mass resolution (m$_{\rm baryon,ini} = 4200\, \Msun$).

The second suite of simulations we use is ``ELVIS on FIRE''.
This suite consists of three simulations, containing two MW/M31-mass galaxies each, wherein the main halos were selected to mimic the relative separation and velocity of the MW-M31 pair in the LG \citep{GK2014,GK2019a,GK2019b}.
ELVIS on FIRE has $\approx 2 \times$ better mass resolution than Latte: the Romeo \& Juliet and Romulus \& Remus simulations have m$_{\rm baryon,ini} = 3500\, \Msun$ and the Thelma \& Louise simulation has m$_{\rm baryon,ini} = 4000\, \Msun$.

We ran all simulations with the upgraded FIRE-2 implementations of fluid dynamics, star formation, and stellar feedback \citep{Hopkins2018}.
FIRE uses a Lagrangian meshless finite-mass (MFM) hydrodynamics code, \textsc{GIZMO} \citep{Hopkins2015}.
\textsc{GIZMO} enables adaptive hydrodynamic gas particle smoothing depending on the density of particles while still conserving mass, energy, and momentum to machine accuracy. 
Gravitational forces are solved using an upgraded version of the $N$-body \textsc{GADGET-3} Tree-PM solver \citep{Springel2005}.

The FIRE-2 methodology includes detailed subgrid models for gas physics, star formation, and stellar feedback.
Gas models used include: a metallicity-dependent treatment of radiative heating and cooling over $10-10^{10}$ K \citep{Hopkins2018}, a cosmic ultraviolet background with early HI reionization ($z_{\rm reion}\sim10$) \citep{FaucherGiguere2009}, and turbulent metal diffusion \citep{Hopkins2016,Su2017,Escala2018}. 
We allow gas that is self-gravitating, Jeans-unstable, cold (T $<10^4$ K), dense ($n>1000$ cm$^{-3}$), and molecular (following \citet{Krumholz2011}) to form stars.
Star particles represent individual stellar populations under the assumption of a Kroupa stellar initial mass function \citep{Kroupa2001}.
Once formed, star particles evolve according to stellar population models from \textsc{STARBURST99} v7.0 \citep{Leitherer1999}.
We model several stellar feedback processes including core-collapse and Type Ia supernovae, continuous stellar mass loss, photoionization, photoelectric heating, and radiation pressure.

For all simulations, we generate cosmological zoom-in initial conditions at $z = 99$ using the \textsc{MUSIC} code \citep{Hahn2011}, and we save 600 snapshots from $z = $ 99 to 0, with typical spacing of $\lesssim$25 Myr.
All simulations assume flat $\lcdm$ cosmologies, with slightly different parameters across the full suite: $h = 0.68 - 0.71$, $\Omega_\Lambda = 0.69 - 0.734$, $\Omega_m = 0.266 - 0.31$, $\Omega_b = 0.0455 - 0.048$, $\sigma_8 = 0.801 - 0.82$, and $n_{\rm s} = 0.961 - 0.97$, broadly consistent with \citet{PlanckCollaboration2018}.

\subsection{Halo finder}\label{halo_finder_subsection}

We use the \textsc{ROCKSTAR} 6D halo finder \citep{Behroozi2013a} to identify dark matter halos and subhaloes in our simulations.
We include a halo in the catalog if its bound mass fraction is $> 0.4$ and if it contains at least 30 dark matter particles within a radius that encloses 200 times the mean matter density, R$_{\rm 200m}$.
We generate a halo catalog for each of the 600 snapshots of each simulation, using only dark matter particles.
The subhaloes that we use in this work (within 300 kpc of their host) are uncontaminated by low-resolution dark matter particles.
We then construct merger trees using \textsc{CONSISTENT-TREES} \citep{Behroozi2013b}.

We describe our post-processing method for assigning star particles to (sub)halos further in \citet{Samuel2020}.
First, we identify all star particles within 0.8 R$_{\rm halo}$ (out to a maximum 30 kpc) of a halo as members of that halo.
Then, we further clean the member star particle sample by selecting those (1) that are within 1.5 times the radius enclosing 90 per cent of the mass of member star particles (R$_{90}$) from both the center-of-mass position of member stars and the dark matter halo center, and (2) with velocities less than twice the velocity dispersion of member star particles ($\sigma_{\rm vel}$) with respect to the center-of-mass velocity of member stars.
We iterate through steps (1) and (2) until the total mass of member star particles ($\Mstar$) converges to within 1 per cent.
Finally, we save halos for analysis that contain at least 6 star particles and that have an average stellar density $> 300\, \Msun \, \mathrm{kpc}^{-3}$.
We performed this post-processing and the remainder of our analysis using the \texttt{GizmoAnalysis} and \texttt{HaloAnalysis} software packages \citep{WetzelHaloAnalysis2020,WetzelGizmoAnalysis2020}.

\subsection{Satellite selection}

Throughout this paper we refer to the central MW/M31-mass galaxies in our simulations as hosts, and their surrounding population of dwarf galaxies within 300 kpc as satellites.
Our host galaxies have stellar masses in the range $\Mstar\sim10^{10-11}\, \Msun$ and dark matter halos in the mass range $\Mtwohm=0.9-1.7 \times10^{12}\, \Msun$.
The eight Latte+m12z simulations contain a single isolated host per simulation.
Each of the three ELVIS on FIRE simulations contains two hosts in a LG-like pair, surrounded by their own distinct satellite populations. 
Thus, we use a total of 14 host-satellite systems to study satellite planes in this work. 
Our fiducial redshift range is $\zzerotwo$ (114 snapshots), giving us a time baseline of $\sim2.4$ Gyr over which to examine the presence of satellite planes at late times in our simulations.
We present our results treating each snapshot as a separate (but not fully independent) realization and stacking snapshots across hosts.
This allows us to mitigate the time-variability and host-to-host scatter in the satellite distribution at small distances from the host, and achieve robust comparisons of simulations and observations.
We also consider a longer time window ($\zzerofive$, 219 snapshots, $\sim5.1$ Gyr) in Section~\ref{time_subsection} in order to examine the lifetimes of planar structures and the coincidence of spatial thinness and kinematic coherence in our simulations.

We consider two ways to select simulated satellite galaxies for comparison to the MW.
Our primary method is to select a fixed number of satellites around each host, by choosing the 14 satellites with highest stellar mass from our simulations, to match the number of observed MW satellites that have $\mstarlowlim$.
We also choose the 15 most massive satellites around hosts for our comparison to M31 (see Section~\ref{m31_subsection} for more details).
Satellites with $\mstarlowlim$ contain $\geq20$ star particles and have peak halo masses of $\mpeaklowlim$ ($\gtrsim 2.3\times10^4$ dark matter particles prior to infall).
Satellite galaxies with $\mstarlowlim$ are also nearly complete in observations \citep[e.g.][]{Koposov2007,Tollerud2008,Walsh2009,Tollerud2014,Martin2016}, so we choose this as our nominal stellar mass limit to select satellites around the MW and M31. 
As an example, at $\zzero$, the satellite with the lowest stellar mass in our fixed-number satellite selection criteria has $\Mstar= 5.6\times10^4\, \Msun$ ($11$ star particles), which is enough to at least indicate the presence of a true satellite, given that it also satisfies the subhalo criteria outlined in Section~\ref{halo_finder_subsection}.

We also consider a stellar mass threshold selection method in Section~\ref{selection_effects_subsection} whereby we require satellites to have $\mstarlowlim$ and maintain the same distance cutoff ($\dlim$).
This selection means that the number of satellites considered around all hosts varies from 10 to 31 in the redshift range $\zzerotwo$.
See \citet{Samuel2020} for more details on the radial distributions and resolution of simulated satellites meeting our criteria, and completeness estimates in the Local Group.
See \citet{GK2019a,GK2019b} for how the stellar mass, velocity dispersion, dynamical mass, and star-formation histories of satellite dwarf galaxies in our simulations all broadly agree with MW and M31 observations, making these simulations compelling to use to examine planarity.

\section{Observations}\label{observations}

We consider all known MW satellite galaxies with $\mstarlowlim$ and $\dlim$, based on the satellite stellar masses and galactocentric distances listed in Table A1 of \citet{GK2019a}.
While we are not confident that our halo finder is able to correctly identify analogues of the Sagittarius dwarf spheroidal (Sgr I) galaxy, given its significant tidal interactions, we include it in our observational sample, because it is a historical member of the MW's satellite plane.
Excluding Sgr I from the MW satellite galaxy sample does not significantly change the resulting spread in the MW's plane metrics, and therefore we achieve essentially the same results in our comparisons to simulations regardless of this choice.
For each observed satellite, we take the sky coordinates and heliocentric distances with uncertainties from \citet{McConnachie2012}.
Furthermore, we include Crater II and Antlia II, which meet our stellar mass and distance criteria as described in \citet{Samuel2020}, and use the positions and uncertainties from their discovery papers \citep{Torrealba2016,Torrealba2018}.
This brings the total number of MW satellites that we consider in this study to 14.
We consider effects of observational incompleteness from the Galactic disc in Section~\ref{obs_completeness}.

We use proper motions from $\Gaia$ Data Release 2 as presented in \citet{Fritz2018}.
We use the larger of the statistical or systematic uncertainties on $\Gaia$ proper motions, which typically is the systematic uncertainties.
We take line-of-sight heliocentric velocities ($\vlos$) for MW satellites and their uncertainties from \citet{Pawlowski2020} and \citet{Fritz2018}, where available.
To supplement this, we use the proper motions and $\vlos$ for the Magellanic Clouds presented in \citet{Kallivayalil2013}, and Antlia II's kinematics come from its discovery paper \citep{Torrealba2018}.

In our analysis of the MW satellite plane, we first sample the heliocentric distances, line-of-sight velocities, and proper motions 1000 times assuming Gaussian distributions on the uncertainties.
We then convert these values to a Cartesian galactocentric coordinate system using Astropy \citep{astropy:2013,astropy:2018}.
We measure planarity on the resulting satellite phase space coordinates in the same way we describe for simulated satellites in Section~\ref{methods}.

We take a different approach to sample M31's satellites.
We impose the same stellar mass limit of $\mstarlowlim$ and 3D distance limit of $\dlim$, but we additionally require that the projected distance from M31 listed in \citet{McConnachie2012} adhere to $\dlimand$, because M31's satellite population is most complete within this range from the Pan-Andromeda Archaeological Survey \citep[PAndAS,][]{McConnachie2009} coverage.
We sample 1000 line-of-sight distances for each satellite, using the posterior distributions published in \citet{Conn2012} where available, and elsewhere assuming Gaussian distributions on distance uncertainties \citep{McConnachie2012,Martin2013a}.
We assume that M32 and NGC205 have the same posterior distance distribution as M31 itself because they are too close to M31 to reliably determine their line-of-sight distances.
The double-peaked posteriors of AndIX and AndXXVII cause the actual number of satellites within $\dlim$ of M31 in each sample to range from 14 to 16, but this is unlikely to cause significant differences in our analysis.
We take the line-of-sight velocities for M31 satellites from \citet{McConnachie2012,Tollerud2012,Collins2013}, and we use them for the 2D kinematic coherence metric described in Section~\ref{methods}.

\begin{figure*}
    \centering
    \includegraphics[width=\textwidth]{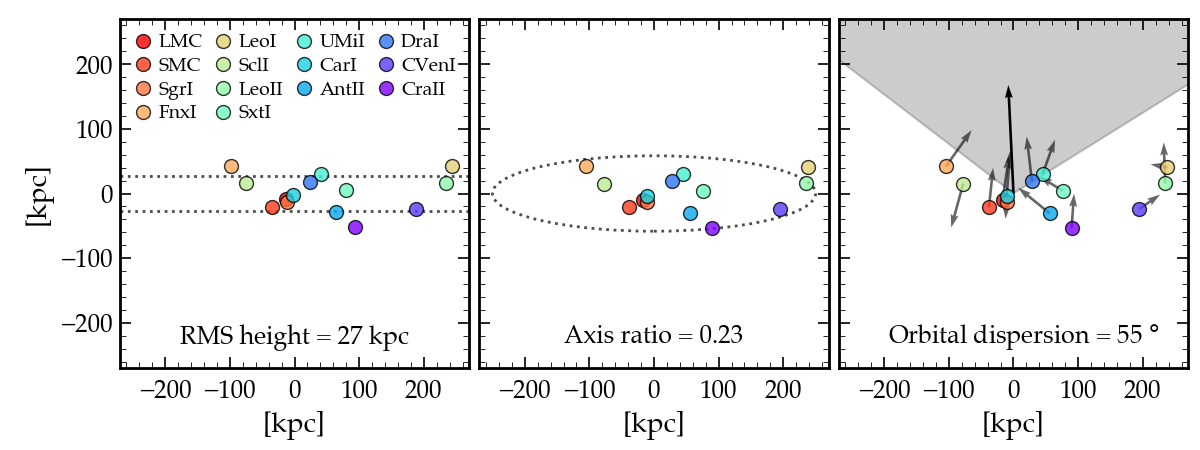}
    \vspace{-7 mm}
    \caption{
    Diagram showing each plane metric that we use, as measured on the 3D positions and velocities of 14 MW satellites ($\mstarlowlim$ and $\dlim$), shown in order of decreasing stellar mass. 
    All planes are centered on the MW and observational uncertainties are neglected here for visual clarity.
    RMS height ($\rmsh$, left) is the root-mean-square distance of satellites from the satellite midplane.
    Axis ratio (middle) is the ratio of the minor-to-major axes ($c/a$) from the moment of inertia tensor of satellite positions.
    The ellipse shown has the same minor-to-major axis ratio as the MW's satellites.
    Orbital pole dispersion ($\orb$, right) is the root-mean-square angle in the range $[0\degree, 360\degree]$ of the angular momentum unit vectors of satellites around their average direction.
    We show each metric in the same projection, to illustrate that the MW's satellite plane is kinematically coherent \textit{within} a spatially thin plane.
    }
    \label{fig:plane_metrics_diagram}
\end{figure*}

\section{Methods}\label{methods}

Figure~\ref{fig:plane_metrics_diagram} is a visual demonstration of how we measure planarity using two spatial metrics and one kinematic metric.
We show these metrics as measured on the MW's 14 satellites with $\mstarlowlim$ and $\dlim$.
For clarity we do not show the effects of observational uncertainties here, which have the largest effect on kinematic coherence, but we do include them in our analysis.
Our planarity metric definitions are based on and consistent with those from e.g., \citet{Cautun2015,Pawlowski2015,Pawlowski2020}.
We require all planes to pass through the center of the host galaxy.
Below, we describe in detail each metric and how we calculated it at each simulation snapshot.

\subsection{Spatial metrics of planarity}\label{spatial_subsection}

We measure the spatial coherence of satellite galaxies in two ways: root-mean-square (RMS) height ($\Delta_{\rm h}$) and minor-to-major axis ratio ($c/a$).
The RMS height of a satellite distribution characterizes the vertical spread of satellites above and below a plane using the RMS component of satellites' 3D positions along the direction normal to a plane according to Equation~\ref{eq1}.
This can be thought of as the thickness or height of the plane.
We randomly generate $10^4$ planes centered on the host galaxy and quote the minimum value amongst these iterations.

\begin{equation}\label{eq1}
    \Delta_{\rm h}=\sqrt{\frac{\sum_{\rm{i}=1}^{\rm N_{sat}}(\uvec{n}_{\perp}\cdot\vec{x}_{\rm i})^2}{\rm N_{sat}}}
\end{equation}

We also use the minor-to-major axis ratio ($c/a$) of the satellite spatial distribution to characterize spatial planes with a dimensionless metric. 
This is the ratio of the square root of the eigenvalues of the inertia tensor corresponding to the minor ($c$) and major ($a$) axes. 
We define a modified moment of inertia tensor treating satellites as unit point masses, weighting each one equally regardless of its stellar or halo mass, so it is a purely geometrical measure of the satellite distribution.
The elements of the 3D inertia tensor are given by Equation~\ref{eq2}.

\begin{equation}\label{eq2}
    I_{ij} = \sum_{k = 1}^{\nsat}\sum_{\alpha = 1}^{3} \delta_{ij}r_{\alpha,k}^2 - r_{\alpha i,k}r_{\alpha j,k}
\end{equation}

We explored a third metric of spatial planarity, enclosing angle, motivated by the desire to mitigate effects of radially concentrated satellite distributions on planarity measurements.
We define enclosing angle as the smallest angle that encompasses the population of satellites, as measured off of the `midplane' of the satellite plane.
Similar to the galactocentric latitude ($b_c$) used in Section~\ref{selection_effects_subsection}, the coordinate origin is placed at the center of the host galaxy. 
Enclosing angle ranges from 0 to 180 degrees by definition, where a measured angle of near 180 degrees indicates an isotropic distribution of satellites.
Similar to the method used for RMS height, in practice we randomly orient planes centered on the host galaxy from which to measure enclosing angle, and find the minimum angle from these iterations.
We found that this metric was significantly noisier over time compared to the other spatial metrics, and often selected a different plane orientation from RMS height and axis ratio, so we do not use it in our final analysis.

\subsection{Kinematic metrics of planarity}\label{kinematic_subsection}

We consider both 3D and 2D measures of orbital kinematic coherence of satellite populations to compare against observed 3D velocities of satellites in the MW, and line-of-sight velocities ($\vlos$) of satellites around M31.
The 3D metric we use is orbital pole dispersion ($\orb$), which describes the alignment of satellite orbital angular momenta relative to the average satellite orbital angular momentum vector for the entire satellite population.
We are not taking into account the magnitude of satellite orbital velocities, so orbital pole dispersion is a measure of purely directional coherence in satellite orbits.
The orbital pole dispersion is defined as the RMS angular distance of the satellites' orbital angular momentum vectors with respect to the population's average orbital angular momentum direction, and is given by Equation~\ref{eq3}.
A system with all satellite orbital angular momenta aligned will have $\orb=0\degree$, while a random, isotropic distribution of satellite velocities has $\orb\sim180\degree$.

\begin{equation}\label{eq3}
    \Delta_{\rm orb}=\sqrt{\frac{\sum_{i=1}^{\rm N_{sat}}[\arccos(\uvec{n}_{\rm orb,avg}\cdot\uvec{n}_{\rm orb,i})]^2}{\rm N_{sat}}}
\end{equation}

To investigate 2D orbital kinematic coherence around M31 we examine whether satellites share the same `sense of orbital direction' around their host galaxy.
We measure this by computing the maximum fraction ($\fmax$) of satellites with opposing (approaching or receding) $\vlos$ on the left and right `sides' of a satellite distribution.
A fraction close to unity indicates a highly coherent system, and a fraction of 0.5 represents a purely isotropic system.
We compute this fraction along $10^3$ randomly generated lines of sight in the simulations, and use the full distribution to compare to M31 as described in Section ~\ref{m31_subsection}.

\subsection{Statistically isotropic realizations of satellite positions and velocities}\label{isotropic_subsection}

To compare the `true' satellite planes (as measured at each snapshot) across different simulations, we quantify the likelihood of measuring thinner or more kinematically coherent planes in a statistically isotropic distribution of satellites.
This is a more general characterization of planarity, independent of the actual values measured for observed systems, that can also address whether satellite planes are statistically significant.
We generate isotropic realizations of satellite positions by randomly generating $\nitertenk$ polar and azimuthal angles for each satellite, keeping their radial distance from the host fixed, following \citet{Cautun2015}.
For isotropic kinematic distributions, we generate random unit velocities (using a similar prescription as for the randomization of angular coordinates) while also randomizing the angular spatial coordinates of each satellite.
We then measure planarity for each of the $\nitertenk$ realizations.
We quantify the significance of a planar alignment by quoting the fraction ($f_{\rm iso}$) of isotropic realizations with smaller values of plane metrics than the true value at each snapshot.
In effect this is the conditional probability of finding a more planar distribution of satellites among the isotropic realizations.
A fraction $f_{\rm iso}\leq0.5$ indicates that the true satellite distribution is more planar than a statistically isotropic distribution of satellites, and we define $f_{\rm iso}\leq0.05$ to mean the true satellite distribution is significantly planar.

\begin{figure*}
    \centering
    \includegraphics[width=\textwidth]{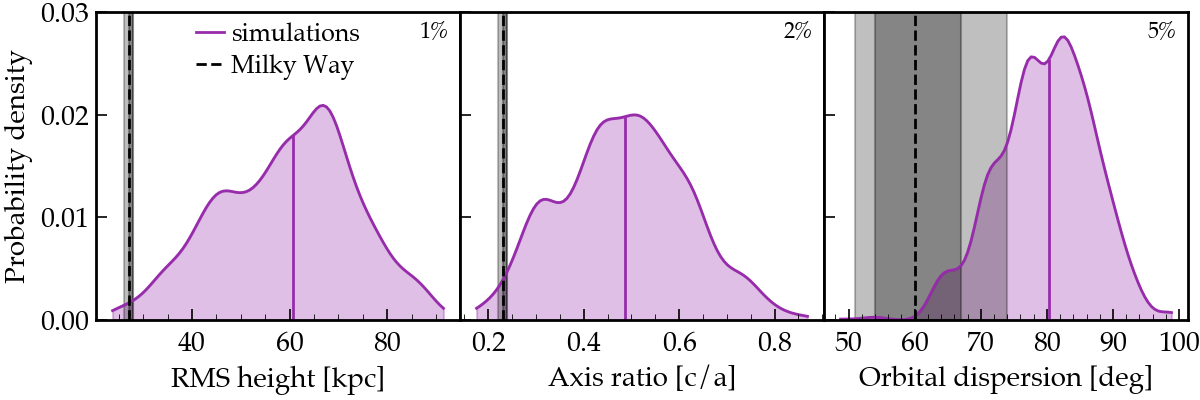}
    \vspace{-6 mm}
    \caption{
    Planarity of simulated satellite galaxies ($\nsatft$ and $\dlim$) around MW/M31-mass hosts compared to the MW's satellite plane.
    We model incompleteness in the simulations by excluding any satellites that lie within $\pm12\degree$ of the plane of the host galaxy's stellar disc.
    We generate KDEs (purple) using 114 snapshots over $\zzerotwo$ for each of the 14 simulated hosts, and the solid vertical colored lines are the distribution medians.
    We show MW observations (black) for 14 satellites with 68 (95) per cent spread from observational uncertainties. 
    The number in the top right of each panel is the per cent of snapshots that are MW-like, which lie at or below the MW upper 68 per cent limit.
    For all metrics we consider, we find some ($1-5$ per cent) snapshots that are at least as planar as the MW, though they are rare.
    }
    \label{fig:mw_completeness_planes}
\end{figure*}

\section{Results}\label{results}

\subsection{Comparisons of simulations and the Local Group}

As we showed in \citet{Samuel2020}, the simulations are a reasonable match to the radial distribution of satellites in the LG as a function of both distance from the host and stellar mass of the satellite.
This provided an important first benchmark of just the 1D radial positions of satellites in our simulation.
We now seek to leverage the full 3D positions and velocities of satellites in our simulations (and around the MW) to characterize satellite planes.
We compare our simulations to observations of LG satellites, leaving comparisons to systems such as other MW/M31 analogues and Centaurus A for future work.
In this section, we make physically rigorous comparisons using mock observations that include disc completeness corrections.
In subsequent sections we further explore selection effects on measured satellite planes and possible physical origins of satellite planes.

\subsubsection{MW-like planes}\label{mw_subsection}

We select the 14 most massive satellites in $\Mstar$ within $\dlim$ to compare planarity in simulations and the 14 MW satellites in our observational sample.
Furthermore, we apply a simple completeness correction for seeing through the MW's disc by first excluding all satellites that lie within a galactocentric latitude of $|b_c|\leq12\degree$ from the host's galactic disc \citep{Pawlowski2018}, and then choosing the 14 most massive satellites from the remaining population.
See Section~\ref{selection_effects_subsection} for an investigation of how disc incompleteness affects planarity metrics.

Figure~\ref{fig:mw_completeness_planes} shows plane metrics for simulated satellites stacking over 114 snapshots spanning $\zzerotwo$, compared to the MW satellite plane.
Spatial plane metrics for the MW are tightly constrained by well-measured 3D positions of MW satellites.
The MW's satellite plane is thinner and more kinematically coherent than most of our simulated satellite systems.
We define MW-like planes as those with plane metrics at or below the one sigma upper limit on the MW's corresponding distribution.
Notably, the MW's plane is significantly spatially flattened compared to the average simulation when measured by RMS height and axis ratio.

While MW-like spatial planes are rare in our simulations, we do identify satellite populations that are as thin as the MW's plane in $1-2$ per cent of our full sample.
We compute each plane metric independently, but we discuss instances of satellite planes that are simultaneously both thin and kinematically coherent in Section~\ref{time_subsection}.
The occurrence of thin planes in $1-2$ per cent of snapshots holds over both our fiducial time baseline of $\zzerotwo\approx2.4$ Gyr (114 snapshots per host, 1,596 snapshots in total) and also over the longer interval $\zzerofive\approx5.1$ Gyr (219 snapshots per host, 3,066 snapshots in total), an indication of the robustness of the measurement.

\begin{figure*}
    \centering
    \includegraphics[width=\textwidth]{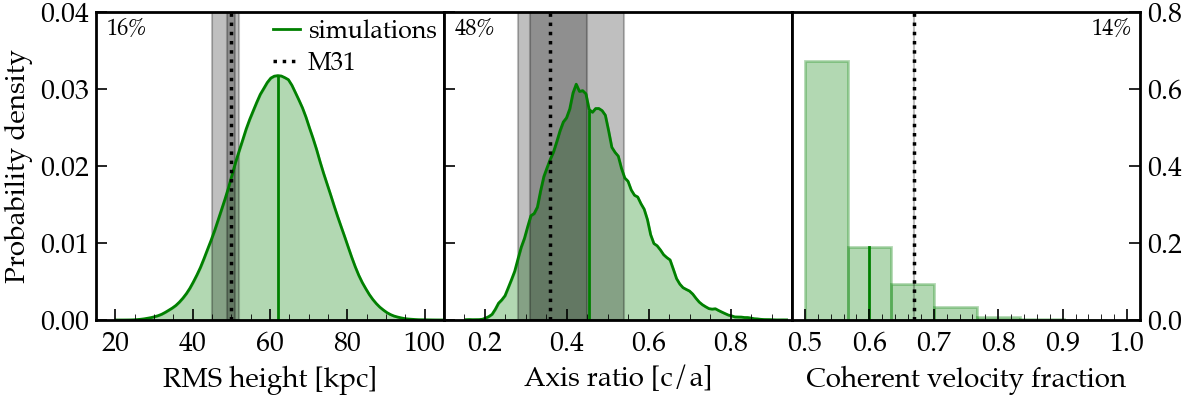}
    \vspace{-6 mm}
    \caption{Planarity of the 15 simulated satellite galaxies with the highest stellar mass within $\dlimand$ of each MW/M31-mass host (consistent with completeness in PAndAS). 
    We generate KDEs (green) using 114 snapshots over $\zzerotwo$ for each of the 14 simulated hosts.
    The solid vertical colored lines are the medians of each distribution.
    We show the M31 data (black) for 16 satellites with 68 (95) per cent spread in plane metrics from line-of-sight (LOS) distance uncertainties. 
    The number in the top of each panel is the per cent of snapshots that are at least as planar as the M31 upper 68 per cent limits.
    Selected in this general way, the simulations are about as planar as M31.
    The average RMS height (left) of simulations is somewhat thicker than M31's satellite population as a whole, but M31 is still within $\sim1\sigma$ of the simulation peak.
    Typical simulation axis ratios (center) are even more similar to M31's satellites.
    In the right panel, more planar snapshots are shown to the right of the M31 value.
    LOS velocity uncertainties are too small to broaden the M31 velocity coherence measurement
    M31's satellites are slightly more kinematically coherent than most simulations, but only by $\sim1\sigma$, consistent with the spatial planarity comparisons.
    }
    \label{fig:m31_planes}
\end{figure*}

The uncertainties in 3D velocities of MW satellites are much larger than the uncertainties in their 3D positions, and this leads to a much wider spread in orbital pole dispersion of the MW compared to the spatial metrics.
However, the MW's satellites still have highly correlated orbits relative to the simulations, with only 5 per cent of the simulations having a plane at least as kinematically coherent as the MW's upper one sigma limit during $\zzerotwo$.
The fraction of the full sample containing these planes actually increases to 8 per cent when measured over $\zzerofive$, likely from the correlated infall of satellites in groups or along filaments at earlier times.
The spread in the MW's orbital pole dispersion is large compared to the spatial metrics, so we also provide the fraction of the simulation sample lying at or below the median MW value, 0.3 per cent.
There are even a few (5) snapshots that extend below the MW distribution.

The MW's satellite kinematics, while rare, do not appear to be extreme outliers compared to our simulations.
This broadly agrees with \citet{Pawlowski2020}, who found that $\sim2-3$ per cent of hosts at $\zzero$ in the IllustrisTNG simulations \citep{Pillepich2018,Nelson2019} have satellites as orbitally aligned as the MW.
However, the comparison between our work and theirs is not one-to-one: they vary the number of satellites included in plane calculations ($\nsat=3-11$) in both simulations and observations in order to account for the ``look elsewhere'' effect (the spurious detection of high significance events from searching a large parameter space), but they find that their conclusions do not vary for any number of plane members greater than three.
The IllustrisTNG simulations they use allow them to analyze a larger number of hosts, in part because they choose to include dark subhaloes as satellites in order to maximize their sample size of hosts with at least 11 satellites.
The larger host sample size comes at the cost of resolution though, with $m_{\rm DM} = 7.5\times10^6 \, \rm{M}_{\odot}$, $m_{\rm baryon} = 1.4\times10^6 \, \rm{M}_{\odot}$, and $\epsilon_{\rm DM, *} = 0.74$ kpc.

In contrast, our planarity metrics are predicated on matching the number of observed satellites ($\nsatft$) and we only have 14 hosts.
Instead, we leverage our time resolution to increase our sample size given that our planes are often transient features (see Section~\ref{time_subsection}).
Our simulations also have order-of-magnitude higher resolution, which may allow planes of satellites to survive that would be disrupted in lower resolution simulations.  This is evidenced by their broad agreement with the MW and M31 in their radial distributions down to $\sim50$ kpc \citep{Samuel2020}.
Our measured plane metrics should be considered upper limits on absolute planarity at each snapshot. 
If we instead varied $\nsat=3-14$, to test for the look-elsewhere effect, we might find even thinner or more coherent planes.
Likewise, our quoted fractions of MW-like planes are upper limits on the incidence of MW-like planarity, as this can only be diminished by accounting for the look-elsewhere effect.
Because we are always choosing a larger number of plane members ($\nsatft$) than used by \citet{Pawlowski2020}, which yields larger values of plane metrics in our case, we compare just the fractions of our samples that are MW-like instead of absolute plane metrics.

As a caveat to these kinematic results, we note that using a slightly different proper motion sample for the observed MW satellites leads to a reduced fraction of snapshots with MW-like kinematic planes.
If we adopt the `best-available' observed proper motions from \citet{Pawlowski2020}, the fraction of snapshots having a plane at least as kinematically coherent as the MW's upper one sigma limit during $\zzerotwo$ decreases to 0.3 per cent (see Figure~\ref{fig:proper_motion_appendix}).
However, this different proper motion data set does not qualitatively change any of our other results in the following sections, and it has no effect on the measured spatial planarity in simulations or observations.
Importantly, we note that the results of Section~\ref{lmc_subsection} still hold: we are more likely to measure a MW-like kinematic plane in the presence of an LMC analogue near first pericenter relative to the general simulation sample.
These caveats are detailed further in Appendix~\ref{proper_motion_appendix}.

As a more rigorous test, we examine the instances of planarity for which simulations are simultaneously spatially thin and kinematically coherent.
We do not find any such simultaneously thin and coherent instances during $\zzerotwo$ in the simulations.
However, looking further back in time to $z=0.5$, we find 10 snapshots that are simultaneously as thin and kinematically coherent as the MW satellites are today.
This amounts to 0.3 per cent of the total sample of snapshots over $\zzerofive$.
This level of simultaneous spatial and kinematic planarity agrees with \citet{Pawlowski2020}, who find that thin and coherent MW-like planes occur in $<0.1$ per cent of IllustrisTNG hosts, than when we examine individual plane metrics.
Notably, the instances of simultaneous planarity in our simulations occur in 2 out of 14 hosts (m12b and m12z).
In both cases, the simultaneous spatial and kinematic planarity occurs around the time of the first pericentric passage of a massive ($\Mstar\geq10^8\, \Msun$) satellite galaxy.
The massive satellite that passes near m12b meets our criteria for being an LMC analogue.
We explore the influence of LMC-like companions further in Section~\ref{lmc_subsection}.

We do not see a significant difference in planarity between satellites of isolated hosts and satellites of hosts in LG-like pairs.
Both the medians and ranges of plane metrics for each host type are essentially the same, so we do not further separate our results by host type.
In Section~\ref{isotropic_subsection}, when we compare true satellites distributions to statistically isotropic distributions, the paired and isolated hosts do not appear systematically different from each other either.
This is consistent with results from \citet{Pawlowski2019}, who reported no significant differences in planarity between dark matter-only simulations of isolated MW-mass halos and paired LG-like halos in the ELVIS simulations \citep{GK2014}.

\subsubsection{M31-like planes}\label{m31_subsection}

For comparison to M31's satellites, we mimic the completeness of PAndAS in our simulations.
We first select all simulated satellites within $\dlim$.
Then, we randomly choose a line of sight from which to observe the simulation, and we select only the satellites that fall within a (2D) projected radius of 150 kpc from the host galaxy.
We choose the 15 satellites with greatest stellar mass that fall within our mock PAndAS-like projection, to match the number of M31 satellites in our observational sample.
We repeat this process along $10^3$ random lines of sight.

In order to meet the 15 satellite criteria, we do not impose a lower limit on the stellar mass of satellites.
At $\zzero$, the lowest mass satellite included in this sample has $\Mstar \approx 1.8 \times 10^4\, \Msun$.
While most simulations easily meet the 15 satellite criteria, there are a few hosts with snapshots that have fewer than 15 luminous satellites within the mock survey area, so we exclude these snapshots.
For example, at $\zzero$, four of the isolated hosts have fewer than 15 satellites selected (as few as 9 satellites) for some lines of sight.
All simulations meet the satellite quota along most lines of sight, and in particular the hosts in LG-like pairs never suffer from this issue.
The results that we achieve with this satellite number selection method are essentially the same as for a stellar mass selection method ($\mstarlowlim$).
We use the full 3D phase space coordinates of these satellites to calculate spatial plane metrics, because the 3D spatial coordinates of each satellite within the coverage of PAndAS are well known.
We calculate planarity metrics along each of $10^3$ lines of sight at each snapshot over $\zzerotwo$ for each simulated host.

Figure~\ref{fig:m31_planes} shows that when considering the 15 most massive satellites, M31-like planes are common in our simulations.
In particular, the axis ratios of simulated satellite systems are typically as planar as the full sample of M31 satellites.
More than 10 per cent of simulations are more planar than M31 for RMS height, so M31 is slightly thinner than our average simulation, but still lies within $\sim1\sigma$ of the simulation median.
Furthermore, throughout $\zzerotwo$ the simulations have many instances of satellite configurations that are simultaneously as spatially thin and kinematically coherent as M31's satellites are under our selection criteria.

Radial (line-of-sight) velocities are currently the only kinematic information available for all of M31's satellites that we consider, so we cannot compute the 3D orbital pole dispersion of them as we did for the MW's satellites.
We quantify kinematic coherence of satellites using $\fmax$, where a larger fraction indicates greater kinematic coherence (see Section~\ref{kinematic_subsection} for details).
As Figure~\ref{fig:m31_planes} shows, 14 per cent of simulations are more kinematically coherent than M31's satellites, though this is still within about $1\sigma$ of the simulation median.
None of our simulations have all satellites sharing the same sense of orbital direction.
\citet{Buck2016} have pointed out that a 2D metric like $\fmax$ likely overestimates the true 3D kinematic coherence, so we may be overestimating the kinematic coherence in both our simulated and observed samples.
The velocity coherence plot (right panel) is shown as a histogram because the underlying distribution is essentially discretely binned.
Because each satellite population contains 15 satellites, the fraction of satellites sharing coherent velocities varies from 0.53 to 1.0 in steps of $\sim0.07$ (see Section~\ref{methods} for calculation details).

We find that the M31 satellite population as a whole is not significantly more planar than our simulations.
This agrees with \citet{Conn2013} who found that M31's overall satellite population is consistent with a statistically isotropic distribution of satellites, though the 15 most-planar of its satellites lie within an exceptionally thin (12 kpc) plane.
While \citet{Buck2015} use a different plane fitting method different from ours (a fixed-height plane), they also recover many instances of satellite planes as thin as the most-planar subset of M31 satellites.

We stress that our comparison to observations is not predicated on selecting the most planar subset of satellites in either simulations or observations.
This is because we prioritize a wholistic view of the planarity of the satellite population as a whole, rather than highly planar subsets of those satellites.
Other than having coherent LOS velocities, which do not unambiguously indicate orbital coherence, the member satellites of M31's plane are not significantly different from non-members, suggesting that they do not have different formation mechanisms or evolutionary histories \citep{Collins2015}.
In addition, sampling the satellite distributions to calculate plane metrics is computationally expensive (see Section~\ref{observations}), and this is made more difficult by finding optimal planes for all satellite combinations.
We defer such an investigation to future work.

For the rest of this work, we do not investigate M31-like planes further.
Instead, we examine MW-like planes, given that completeness is more certain out to the virial radius, and precise 3D velocities of MW satellites are available.
The availability of 3D velocities of MW satellites provides a more realistic metric of kinematic coherence.

\begin{figure*}
    \begin{multicols}{2}
	\includegraphics[width=0.45\textwidth]{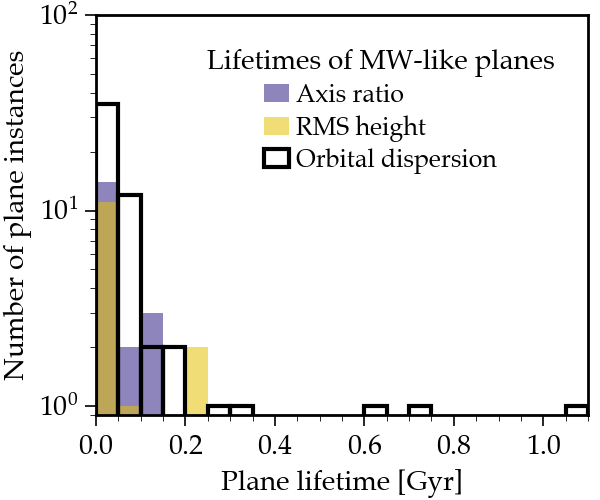}\par
	\includegraphics[width=0.45\textwidth]{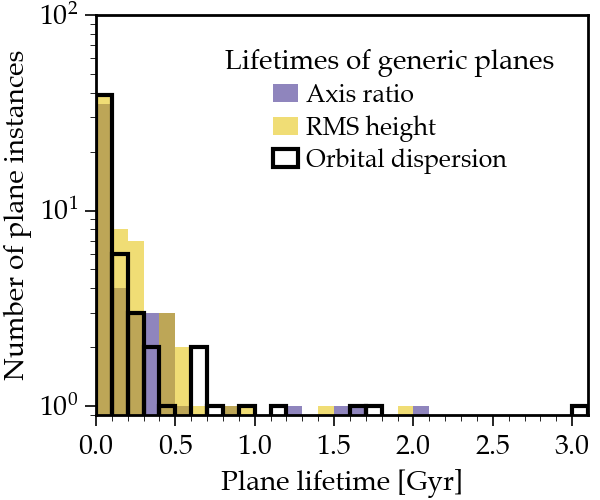}\par
	\end{multicols}
	\vspace{-6 mm}
    \caption{
    Satellite plane lifetimes measured over $\zzerofive$ (219 snapshots per host, $\sim25$ Myr spacing) for the 14 satellites with the greatest $\Mstar$ within $\dlim$. 
    We define lifetimes independently for each plane metric.
    \textit{Left:} MW-like planes are those that have plane metrics at or below the MW upper one sigma limits.
    We have applied the same completeness correction for seeing through the disc as in Section~\ref{mw_subsection}. 
    Such planes are rare and short-lived, with most lasting $<0.5$ Gyr and none surviving for longer than $\sim$1 Gyr. 
    Two out of the three instances of MW-like planes lasting $>$500 Myr occur in hosts that experience a pericenter passage of an LMC-like satellite.
    \textit{Right:} Generic planes are any flattened or kinematically coherent systems whose plane metrics fall below the lower 68 percent limits of our simulations shown in red in Figure~\ref{fig:combined_selections_planes}. 
    Generic planes are also typically short-lived and many last for only a single snapshot. 
    Half of the hosts have an instance of a generic plane that last $>1$ Gyr, and two of those experience an LMC-like pericenter passage.
    While some generic planes live for a few Gyr, those planes are not typically simultaneously spatially thin and kinematically coherent.}
    \label{fig:plane_lifetimes}
\end{figure*}

\subsection{Statistical significance and lifetimes of planes}

\subsubsection{Statistical significance of planes}\label{statistical_subsection}

We now move from absolute metrics of planarity to a more general investigation of planarity, that does not rely on MW or M31 observations to establish what constitutes a planar configuration.
We characterize the statistical significance of satellite planes in our simulations by randomizing the positions and velocities of satellites in order to form a statistically isotropic distribution as a control sample (see Section~\ref{isotropic_subsection} for how we set this up).
By generating $\nitertenk$ isotropic iterations and acquiring plane metrics from them, we create a bank of plane metrics that one might expect to measure if the distribution is statistically isotropic.
This isotropic bank is used to compute plane significance by calculating the fraction ($f_{\rm iso}$) of isotropic iterations that are \textit{more} planar than the true measured value at each snapshot.
In effect, this provides an estimate of the probability of finding a thinner or more coherent plane in a random distribution of satellites.
Small fractions ($f_{\rm iso}\leq0.05$) indicate a rare plane with high significance, while larger fractions ($f_{\rm iso}\geq0.5$) show that the measured plane is consistent with an isotropic distribution of satellites.

We distinguish between two different measures of plane statistical significance: conditional probability and marginalized probability (following \citet{Cautun2015}).
Marginalized probability refers to the significance of a system's planarity relative to an ensemble of planarity measurements on that system where the number of satellites considered is allowed to vary from the minimum number of points needed to define a plane (3) to some maximum.
We concentrate our analysis on conditional probability, because it represents the significance of a system's planarity given a certain set of constraints (such as completeness or total number of satellites).
We calculate the significance of planes on simulations across $\zzerotwo$, and on the observed positions and velocities of MW satellites.
Again, for the simulations, we remove satellites obscured by the host disc at $|b_c|\leq12\degree$.
This is the same selection that we used in Figure~\ref{fig:mw_completeness_planes}.

By these simple metrics, and without correcting for selection or the look-elsewhere effect, the MW's plane is highly significant relative to a statistically isotropic distribution.
Less than one per cent of the MW's isotropic realizations of its satellites have a thinner plane ($f_{\rm iso}=0.003$ for RMS height or axis ratio), or a more kinematically coherent plane ($f_{\rm iso}=0.005$ for orbital pole dispersion).
In comparison, many of our simulation snapshots have median $f_{\rm iso}\gtrsim0.5$, indicating that they are broadly consistent with and have no meaningful degree of planarity relative to a statistically isotropic distribution of satellites.
See Appendix~\ref{isotropic_appendix} for a visual representation of $f_{\rm iso}$ for each host during $\zzerotwo$.
About half of both the isolated and paired hosts have median $f_{\rm iso}<0.5$, and this similarity indicates that the paired host environment does not significantly enhance the statistical significance of satellite planes.
About half of the hosts have $\sim5-10$ per cent of their snapshots with $f_{\rm iso}<0.05$, indicating significant spatial planes for these particular snapshots.

Only 3 out of the 14 hosts have significant kinematic coherence relative to a statistically isotropic distribution of satellite velocities, consistent with previous studies \citep[e.g.][]{Metz2008,Pawlowski2014,Ahmed2017,Pawlowski2020}.
Notably, none of the hosts have satellites that are simultaneously highly spatially significant ($f_{\rm iso}<0.05$) and highly kinematically significant relative to a statistically isotropic distribution at any snapshot during $\zzerotwo$.
In general, hosts that with small ($<0.25$) median $f_{\rm iso}$ for spatial planarity metrics do not have correspondingly small $f_{\rm iso}$ for kinematic coherence (orbital pole dispersion), and vice versa.
While our simulations contain instances of planes that are simultaneously as spatially thin and kinematically coherent at the MW in an absolute sense (by directly comparing plane metrics), the planes found in our simulations are not as significant relative to a statistically isotropic distribution.

\subsubsection{Lifetimes of planes}\label{time_subsection}

Thus far we have focused our analysis on the spatial and kinematic coherence of satellite galaxies in our simulations over $\zzerotwo$ ($\sim$2.4 Gyr).
In this section, we seek to understand if the satellite planes we find are long-lived and stable, or merely transient configurations, across $\zzerofive$ ($\sim5.1$ Gyr, 219 snapshots).
This longer time baseline allows us to examine the time evolution of satellite plane structures as satellites make multiple orbits around their host.
A satellite in the inner regions of its host's halo may complete an orbit in under 1 Gyr.
Satellites in the most outer regions of the host halo take $\sim$3-4 Gyr to undergo a complete orbit.
We consider a plane to be ``long-lived'' if it persists for $\geq1$ Gyr, lasting for at least one satellite orbital timescale in the inner halo.
We deem any planar configurations lasting $<1$ Gyr to be ``short-lived'', and we consider those lasting $<500$ Myr to be ``transient'' alignments that do not indicate coherence amongst satellite orbits because they are so short.

We examine the distribution of plane lifetimes over $\zzerofive$ separately for MW-like planes and generically flattened satellite systems.
We define MW-like planes as those with plane metric values at or below the upper 68 per cent limits on MW values: RMS height $\leq28$ kpc, axis ratio $\leq0.24$, or orbital pole dispersion $\leq67\degree$.
We measure MW-like plane lifetimes on the same simulation data in Figure~\ref{fig:mw_completeness_planes}, which includes a correction for seeing through the host disc.
`Generically' flattened means having plane metric values: RMS height $\leq48$ kpc, axis ratio $\leq0.39$, or orbital pole dispersion $\leq71\degree$, defined by the lower 68 per cent limit on simulation plane metrics during $\zzerotwo$.
We measure generic planes on the simulation data presented in Figure~\ref{fig:combined_selections_planes}, which selects the 14 most massive satellites in stellar mass but does not include a correction for seeing through the host disc.
We measure plane lifetimes ($\Delta t_{\rm plane}$) as the amount of time that a system spends consecutively at or below these plane metric thresholds.
Whether a satellite system is planar for only a single snapshot ($\lesssim25$ Myr) or many consecutive snapshots, we count it as a single instance of planarity.

Figure~\ref{fig:plane_lifetimes} shows that for both MW-like and generic planes, most planar instances are transient alignments and many last for just one snapshot ($\Delta t_{\rm plane} < 25\, \rm{Myr}$).
There are 348 snapshots with MW-like planes in our simulations across all hosts over $\zzerofive$ (219 snapshots per host, 3,066 in total) and amongst all three 3D plane metrics.
Out of the total 89 separate instances of MW-like planes, most (56) are in kinematic coherence and only one of them lasts for $\gtrsim1 \text{Gyr}$.
This only occurs for one host, m12b, which also happens to experience a close passage of an LMC-like satellite during that time, that we discuss further in Section~\ref{lmc_subsection}.
There are 1,796 snapshots and 177 separate instances of generically flattened planes in our simulations.
Almost all generic planes last $<1$ Gyr, with only a small fraction ($<10$ per cent) of separate instances extending up to 3 Gyr.
One host, m12f, has a generic kinematic plane lasting 3 Gyr and also experiences an LMC-like passage during this time.
We conclude that satellite planes in our simulations, regardless of exact plane metric, are typically transient alignments that do not indicate a long-lived orbiting satellite structure, though the presence of LMC-like satellites can lead to longer-lived planes.

We also examine our simulations for instances of satellite configurations that are simultaneously spatially thin and kinematically coherent.
We use the same plane metric thresholds as above to look at how often a satellite system meets the kinematic threshold and at least one of the spatial thresholds at the same snapshot.
We do not find any instances of simultaneously thin and coherent MW-like planes over $\zzerotwo$ using either our fiducial selection method ($\nsatft$ and $\dlim$) or combining that with a completeness correction due to seeing through the host's galactic disc.
However, there are several instances of coincident thinness and coherence over $\zzerofive$, especially when we apply a completeness correction for seeing through the host disc.
In particular, m12b and m12r have up to 13 snapshots ($\sim325$ Myr) of simultaneous spatial and kinematic planarity over $\zzerofive$.
We also consider the coincidence of generic planes, and find that m12b, m12r, and m12f all have snapshots with simultaneous spatial and kinematic planes even without implementing a completeness correction for the host disc.
Both m12f and m12b have LMC satellite analogues during this time, as we discuss in Section~\ref{lmc_subsection}.
Interestingly, none of our hosts in LG-like pairs exhibit simultaneous planarity, reinforcing the result that LG-like host environments are not more likely to have satellite planes.

\citet{Shao2019} looked at plane lifetimes in the EAGLE simulations.
They considered both a different sample size ($\nsat=11$) and a longer time baseline ($z\approx0-2\approx10.5$ Gyr).
This leads them to identify thinner planes in an absolute sense, because fewer satellites create a thinner plane.
This time window may also catch some MW-like hosts as they are still being formed by mergers of smaller galaxies and before they have been able to form most of their stellar mass \citep[e.g.][]{Santistevan2020}.
However, they too found that most instances of MW-like spatially thin planes were short-lived ($<1$ Gyr), but some systems remain orbitally coherent for upwards of 4 Gyr.
Though we do not find such long-lived kinematic planes in our sample, this generally agrees with our findings.

\citet{Fernando2017,Fernando2018} examined the stability of M31-like planes in idealized simulations, and found that most planes are short-lived and plane stability is highly sensitive to initial satellite phase space coordinates, plane alignment with the host halo, and subhalo abundance.
The authors found that satellites moving perpendicular to the plane, misalignment of the plane with the halo axes, and increased subhalo abundance all generally caused planes to disrupt within $\leq3$ Gyr.
While they demonstrated this within idealized simulations and specifically for comparison to the M31 plane, their modeling approach was general enough to compare to our plane lifetime results, where we find similarly short lifetimes for generic planes in cosmological simulations.
This might lead one to conclude that the MW's plane is short-lived. 
However, we note that two of the three instances of MW-like planes with longer lifetimes in our simulations (based on orbital pole dispersion) have something else in common, the presence of an LMC-like satellite.
Such a massive satellite near pericenter that has brought with it its own satellites may contribute to a longer plane lifetime, so the MW's plane may not be as short-lived as the majority of our simulated MW-like planes.
This is discussed further in Section~\ref{lmc_subsection}.

\subsection{Selection effects on measured planarity}

\subsubsection{Observational incompleteness from the host disc}\label{obs_completeness}

In our analysis of MW-like planarity thus far, we have applied a fixed obscuration correction for seeing through the host disc, masking out everything that lies within $|b_c|\leq12\degree$ (where $c$ indicates a galactocentric coordinate system).
We now analyze how the relative incidence of MW-like planes changes as a function of how much of the sky is obscured by the host's disc.
We vary the region obscured by the galactic discs of simulated hosts from $b_c=0\degree$ (completely unobscured) to $|b_c|\leq45\degree$ (majority obscured) in increments of $\Delta|b_c|=3\degree$.
For each obscured region we select the 14 most massive satellites in $\Mstar$ within $\dlim$ of a host to use in the plane sample.
We define the relative incidence of MW-like planes as follows: we compute the fraction of snapshots with MW-like planes for each obscured region, and normalize it to the unobscured ($|b_c|=0\degree$) fraction of snapshots with MW-like planes.
We repeat this process for each plane metric individually.
However, for $|b_c|\geq30\degree$ there are typically fewer than 14 luminous satellites in the unobscured region and near $|b_c|\sim40\degree$ there are only about 10 satellites available on average, so we cannot draw strong conclusions about completeness effects in those limits.

\begin{figure}
	\includegraphics[width=\columnwidth]{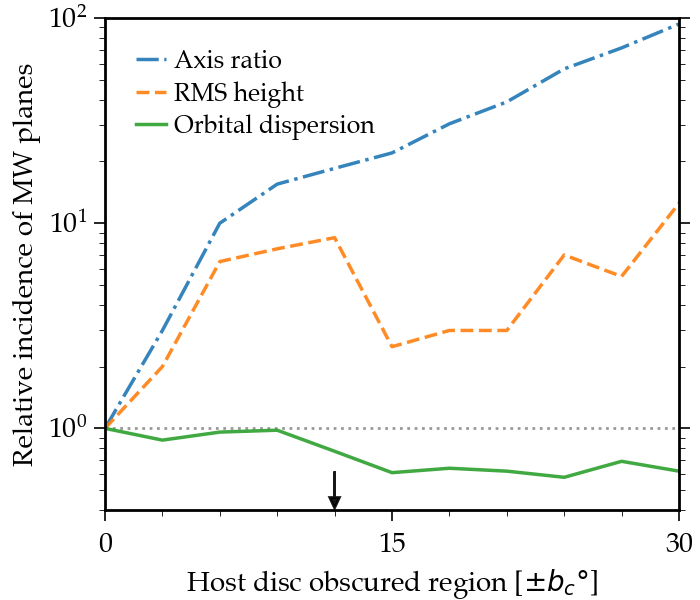}
	\vspace{-5 mm}
    \caption{
    Effects of disc incompleteness on measured planarity.
    We define the relative incidence of MW planes as the fraction of snapshots during $\zzerotwo$ with MW-like planes normalized to the true or unobscured fraction ($|b_c|=0\degree$).
    We select the 14 most massive satellites in $\Mstar$ within $\dlim$ of each host, but for $|b_c|\geq30\degree$ there are usually fewer than 14 luminous satellites available.
    The horizontal line represents consistency with the unobscured fraction. 
    The arrow shows the fiducial obscuration that we adopt for MW-like planes, $|b_c|=12\degree$.
    Spatial planarity ($c/a \leq 0.24$, $\rmsh\leq 28$ kpc) is much more affected by host disc obscuration than kinematic planarity.
    Spatial planarity jumps an order of magnitude between $|b_c|=0\degree$ and $|b_c|\sim10\degree$.
    Kinematic planarity ($\orb\leq67\degree$) is slightly diminished by host disc obscuration.
    At $|b_c|=12\degree$, we are $8.5-18.5\times$ more likely to measure a MW-like spatial plane and $1.3\times$ less likely to measure a MW-like kinematic plane.
    As expected, when nearly half of the sky is obscured spatial planarity is highly likely to be measured.}
    \label{fig:host_disc}
\end{figure}

Figure~\ref{fig:host_disc} shows the incidence of MW-like planes, measured independently for each metric, as a function of disc obscuration angle.
We find that such incompleteness artificially boosts the fraction of snapshots with MW-like spatial planes for any value of $|b_c| > 0$.
In particular, near the fiducial obscuration we adopt for MW-like planes in previous sections ($|b_c|=12\degree$), the incidence of MW-like planes is increased by about an order of magnitude.
For $|b_c|\lesssim40\degree$, disc obscuration has a much smaller and opposite effect on kinematic planarity compared to spatial planarity; MW-like kinematic planes tend to be somewhat washed out by incompleteness.
Near our fiducial obscuration for the MW, the relative incidence of MW kinematic planes is about 0.77.
As expected, disc obscuration has the largest effect on planarity when $|b_c|\sim45\degree$, where so much of the sky is obscured that any detected satellites would appear to be in a plane purely due to incompleteness.

Our results show that observational incompleteness from the host disc can have a strong effect on measured spatial planarity.
If the MW's satellite population is incomplete from seeing through the Galactic disc at our fiducial level, then MW observations may be overestimating the spatial planarity of MW satellites by a factor of $\sim10-20$.
To a much lesser degree, MW observations may underestimate the kinematic coherence of satellites by a factor of $\sim1-2$.
Because this incompleteness may bias our analysis of the underlying causes of satellite planes, we only use a host disc correction when comparing directly to MW observations in Sections~\ref{mw_subsection} -- ~\ref{time_subsection} and ~\ref{lmc_subsection}.
For the remainder of this paper, we do not include a host disc correction.

\begin{figure*}
    \centering
    \includegraphics[width=\textwidth]{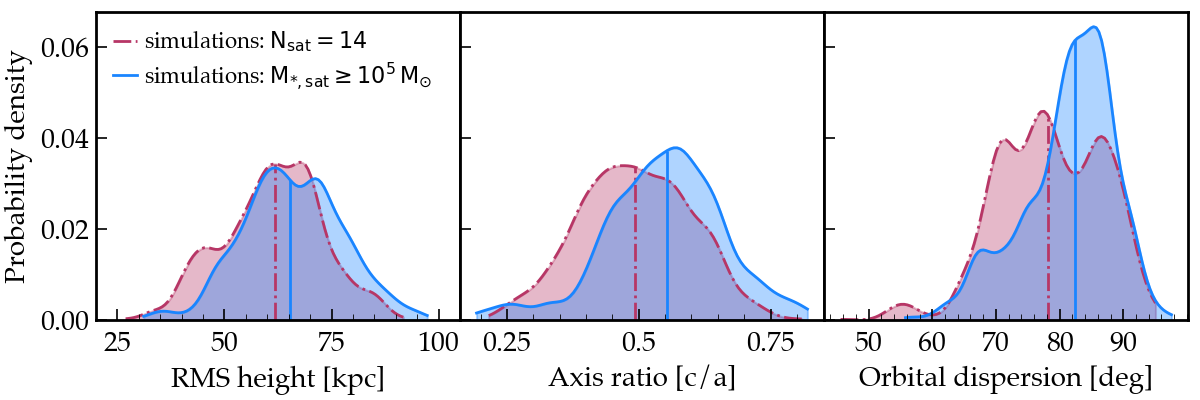}
    \vspace{-6 mm}
    \caption{
    Planarity of simulated satellite galaxies ($\dlim$) selected using a fixed number method versus a stellar mass threshold.
    Note that we do not include a correction for completeness due to seeing through the host galactic disc here.
    We generate KDEs using 114 snapshots over $\zzerotwo$ for each of the 14 simulated hosts.
    The vertical lines are the medians for each distribution.
    The red distributions are the 14 most massive satellites in stellar mass, while the blue distributions are all satellite galaxies with $\mstarlowlim$.
    Thin and coherent planes are rare in the simulations using these particular selections and time baseline, but using the number selection for satellites yields lower (more planar) metrics because the stellar mass selection allows for many more satellites to be included ($\nsat=10-31$).}
    \label{fig:combined_selections_planes}
\end{figure*}

\subsubsection{Method of selecting simulated satellites}\label{selection_effects_subsection}

We also explore how using a fixed number selection for satellites compared to using a stellar mass threshold affects planarity measurements.
Our primary method of satellite selection throughout this work is to choose the 14 most massive satellites by rank-ordering them in stellar mass, because the number of satellites in a sample strongly correlates with the measured planarity \citep[e.g.][]{Pawlowski2019}.
In terms of observational completeness and resolution in simulations, another way to select satellites may be to impose a simple stellar mass threshold.
So we test our fixed number selection against a stellar mass threshold method: $\mstarlowlim$ and $\dlim$.
However, this leads to a range of numbers of satellites selected around each host, which makes it difficult to compare plane metrics across simulations and observations.
The total number of satellites with $\mstarlowlim$ and $\dlim$ per host varies from 10-31 during $\zzerotwo$ in our simulations.

Figure~\ref{fig:combined_selections_planes} shows that planes with $\nsatft$ tend to be both thinner and more kinematically coherent than planes with $\mstarlowlim$, because while some $\mstarlowlim$ satellite populations have $\nsat<14$, more actually have $\nsat>14$.
One consequence is that when using the $\Mstar$ selection the simulations never reach the MW's RMS height (27 kpc) during $\zzerotwo$, but the fixed number selection does.
The small bump in the $\nsatft$ orbital pole dispersion distribution is from a single host, m12f, during the snapshots following a close passage of an LMC-like satellite.
We discuss effects of such an LMC-like companion further in the following section.

This selection exercise highlights an important aspect of the satellite plane problem: many of the conclusions drawn about the nature of satellite planes are sensitive to satellite selection method, likely because of underlying sensitivity to the number of satellites in the sample.
Had we used the stellar mass threshold as our fiducial selection method in previous sections, we would have found more evidence for tension between simulations and observations, but deciding whether that tension is cosmologically significant is hampered by the sensitivity of plane metrics to both incompleteness and sample selection.

\subsection{Exploring physical explanations of planes}

\subsubsection{Influence of an LMC-like satellite}\label{lmc_subsection}

The presence of a massive satellite galaxy near pericenter, like the Large Magellanic Cloud (LMC), has been suggested as a possible explanation for the dynamical origin of the MW's satellite plane \citep{Li2008,DOnghia2008}, from the accretion of multiple satellites in a group with the LMC \citep[e.g.,][]{Wetzel2015b,Deason2015,Jethwa2016,Sales2017,Jahn2019}.
We seek to determine whether or not the presence of an LMC-like companion has an effect on the planarity in simulations.
We compare planarity metrics measured on systems experiencing an LMC-like passage to those without an LMC-like passage.
We identify pericentric passages of four LMC-mass analogues in our simulations based on the following selection criteria:

\begin{enumerate}[(i)]
    \item $t_{\rm peri} > 7.5$ Gyr ($z < 0.7$)
    \item M$_{\rm sub,peak} > 4\times10^{10}\, \Msun$ and $\Mstar > 10^{9}\, \Msun$
    \item $d_{\rm{peri}} < 50$ kpc
    \item The satellite is at its first pericentric passage.
\end{enumerate}

This broad time window allows us to capture a larger number LMC-like passages, which tend to be rare as we have defined them.
The minimum mass is consistent with measurements of the LMC's mass \citep{Saha2010}, and the maximum pericenter distance reflects the measured distance and orbit of the LMC \citep{Freedman2001,Besla2007,Kallivayalil2013}.
Table~\ref{tab:LMC-table} lists the four hosts in our simulations with LMC satellite analogues that meet these criteria, all of which are from simulations of isolated MW-like hosts rather than paired/LG-like hosts.
We emphasize that these satellites are not the only sufficiently massive satellites in the simulations, but that they are the only instances that satisfy all our LMC analogue criteria simultaneously.

To compare planarity during LMC-like passages and otherwise, we first select all snapshots within $\pm5$ snapshots (a time window of $\sim250$ Myr) of the LMC-like pericenter passage in each of the four simulations containing an LMC analogue.
This gives us a total of 44 snapshots that we classify as occurring close enough to an LMC analogue pericenter to exhibit any dynamical effects of group infall.
We compare plane metrics from those snapshots to plane metrics measured on all other simulations (excluding the four hosts with LMC analogues) up to the earliest snapshot included in the LMC sample ($z\sim0-0.7$, 247 snapshots per host).
We apply our fiducial disc obscuration correction, masking out all satellites within $|b_c|\leq12\degree$ of the hosts' galactic discs in our simulations.
To calculate plane metrics we select $\nsatft$ of the most massive satellites ranked by stellar mass.

\begin{table*}
\centering
\caption{
Properties of the LMC satellite analogues at their first pericentric passage about their MW/M31-mass host in our FIRE-2 simulations. We select satellites with M$_{\rm sub,peak} > 4 \times 10^{10}\, \Msun$ and $\Mstar > 10^{9}\, \Msun$ that have their first pericenter after 7.5 Gyr ($z < 0.7$) and within 50 kpc of their host.
}
\begin{tabular}{lllllll}
	\hline
	Host & 
	M$_{\rm sub,bound}$ [$10^{10}\, \Msun$] & 
	M$_{\rm sub,peak}$ [$10^{11}\, \Msun$] &
	$\Mstar$ [$10^{9}\, \Msun$] & 
	$t_{\rm peri}$ [Gyr] & 
	$z_{\rm peri}$ & 
	$d_{\rm peri}$ [kpc] \\
	\hline
	m12b    & 12.0 & 2.1 & 7.1 &  8.8 & 0.49 & 38 \\
	m12c    &  5.1 & 1.6 & 1.2 & 12.9 & 0.07 & 18 \\
	m12f    &  6.0 & 1.5 & 2.6 & 10.8 & 0.26 & 36 \\
    m12w    &  4.9 & 0.8 & 1.3 &  8.0 & 0.59 &  8 \\
	\hline
\end{tabular}
\label{tab:LMC-table}
\end{table*}

\begin{figure*}
    \centering
    \includegraphics[width=\textwidth]{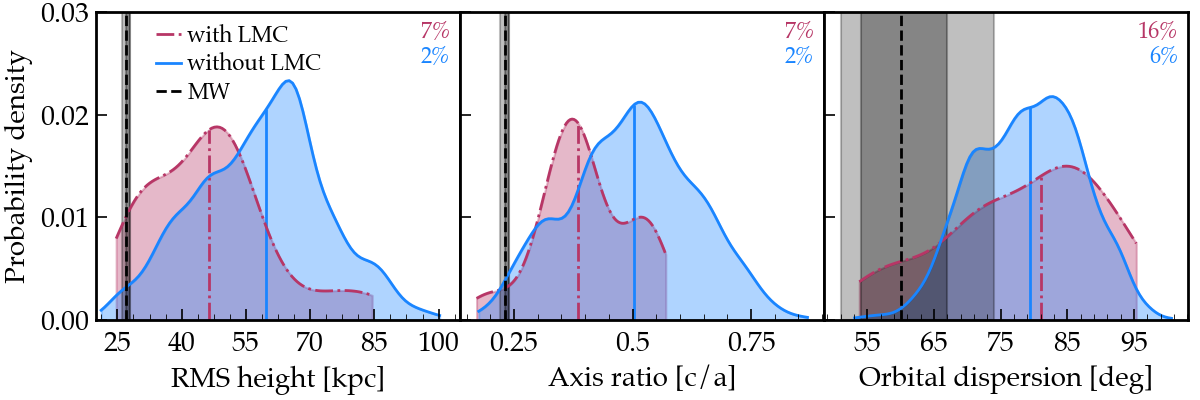}
    \vspace{-7 mm}
    \caption{
    Planarity of satellites of hosts experiencing a first pericenter passage of an LMC satellite analogue (red) compared to all other hosts without an LMC-like passage (blue).
    We rank order satellites by stellar mass and choose the 14 most massive around each host within $\dlim$.
    We select snapshots within $\pm125$ Myr of LMC-like passages that occur during $z\sim0-0.7$.
    Only 4 hosts have such LMC-like passages (see Table~\ref{tab:LMC-table}).
    Vertical colored lines are the medians of the simulation distributions.
    MW planarity values are the vertical black lines and shaded regions.
    We apply a disc obscuration correction and omit satellites within $|b_c|\leq12\degree$.
    LMC passages push towards $\sim20$ per cent lower plane metric medians and smaller ranges of spatial planarity metrics.
    MW-like planes are $\sim2-3$ times more likely to be measured during an LMC-like passage.
    }
    \label{fig:lmc_passages}
\end{figure*}

Figure~\ref{fig:lmc_passages} summarizes our results for the planarity of satellites during an LMC analogue pericenter passage compared to all other satellite systems during $z\sim0-0.7$.
In general, the presence of an LMC analogue leads to thinner and more kinematically coherent satellite planes on average.
The presence of an LMC analogue shrinks the range of spatial plane metric values and slightly shifts them towards smaller (thinner) values.
In particular, the range of axis ratios is much smaller in the presence of an LMC analogue.
The right panel of Figure~\ref{fig:lmc_passages} also shows that the presence of an LMC analogue is correlated with more of the simulation distribution having tighter orbital alignment of satellites.
For all three metrics, we are $\sim2-3$ times more likely to measure a MW-like plane during an LMC pericentric passage compared to the general simulation sample.
This result persists if we widen our time window to $\pm10$ snapshots ($\sim500$ Myr).
The enhancement in the fraction of snapshots with MW-like spatial planes and an LMC near pericenter washes out for time windows larger than $\sim500$ Myr, but the enhancement for MW-like kinematic planes persists in even the largest time window ($\pm40$ snapshots or $\sim2$ Gyr) that we tested.
If we instead use the `best-available' proper motion data set from \citet{Pawlowski2020}, the increased likelihood of measuring a MW-like kinematic plane is actually strengthened, because only 1 per cent of the general snapshot sample during $z\sim0-0.7$ is MW-like versus 11 per cent near LMC analogue pericenters (see Appendix~\ref{proper_motion_appendix}).
Thus, the presence of the LMC on first infall may contribute significantly to the thin and (even more so) kinematically coherent satellite plane around the MW.

We also consider the time evolution of planarity both before and after LMC analogue pericentric passages.
For RMS height and orbital pole dispersion in particular, the hosts begin to experience downward trends in these metrics just before or at the time that the LMC analogue crosses within $R_{\rm 200m}$ of the MW-mass host, reaching minimum values up to few hundred Myr after the LMC analogue's pericentric passage.
As an additional comparison of simulated and observed MW satellite kinematics, we calculate the velocity anisotropy parameter, $\beta$, following \citet{Cautun2017}.
We measure $\beta=-1.35\pm0.2$ for our sample of 14 observed MW satellites, indicating a preference for circular orbits.
The distribution of $\beta$ for the simulations experiencing LMC analogue pericenter passages has a longer and more prominent tail towards more negative values of $\beta$ (more circular orbits), as well as a lower median value than the general simulation sample.
Thus, our analysis of $\beta$ also suggests that the presence of the LMC may increase the likelihood of measuring MW-like satellite kinematics.
While more circular orbits could conceivably lead to a more stable satellite plane, \citet{Cautun2017} found little evidence for a correlation between aligned orbital poles and circularity of satellite orbits.
We leave a full dynamical analysis of the LMC's influence on planarity for future work.

We find that the main reason for enhanced planarity in systems with LMC analogues is that the LMC analogues bring satellites with them that are counted in the plane sample, and because it is only at first pericenter there has not been enough time for the LMC and its satellites to dissociate from each other \citep[e.g.,][]{Deason2015}.
The four LMC analogues each bring in $2-4$ satellites with $\mstarlowlim$, consistent with the results presented in \citet{Jahn2019} for both likely satellites of the LMC and FIRE-2 simulation predictions for satellites of LMC-mass hosts.
Of the $2-4$ LMC analogue satellites, $1-3$ of them are counted toward the $\nsatft$ satellites in the plane sample.
The host with the most planar configuration that we find (m12b), which also has instances of simultaneous spatial and kinematic planarity, brings in four satellites with $\mstarlowlim$ and three of these (plus the LMC analogue itself) are counted in the plane calculations.
This means that the LMC analogue and its satellites account for $\sim30$ per cent of the plane sample for m12b, so spatial and kinematic coherence of the LMC subgroup can easily drive the measured plane metrics to lower values.

\citet{Shao2018} examined whether anisotropic accretion or group accretion could explain the formation of satellite planes.
They ultimately found that most massive satellites were singly accreted, and that anisotropic accretion rather than group accretion correlated more with planarity.
In light of this we test for whether planarity correlates in general with group accretion and average infall times.
We find that overall most of the satellites in our sample were either singly accreted or accreted as groups of two.
We do not find a strong correlation between such group accretion and planarity in our full sample.
So we stress that our key result is that only a sufficiently massive LMC-like satellite near first pericenter shows a clear sign of enhancing planarity.
Two hosts with LMC analogues (m12f and m12b) experience extended periods of planarity ranging from $0.7-3$ Gyr (Section~\ref{time_subsection}).
We conclude that LMC-like satellites contribute significantly to satellite planes with moderate lifetimes ($\sim1-2$ Gyr), but that they are unlikely to have a permanent effect on the satellite distribution on longer timescales.

\begin{figure*}
    \centering
    \includegraphics[width=\textwidth]{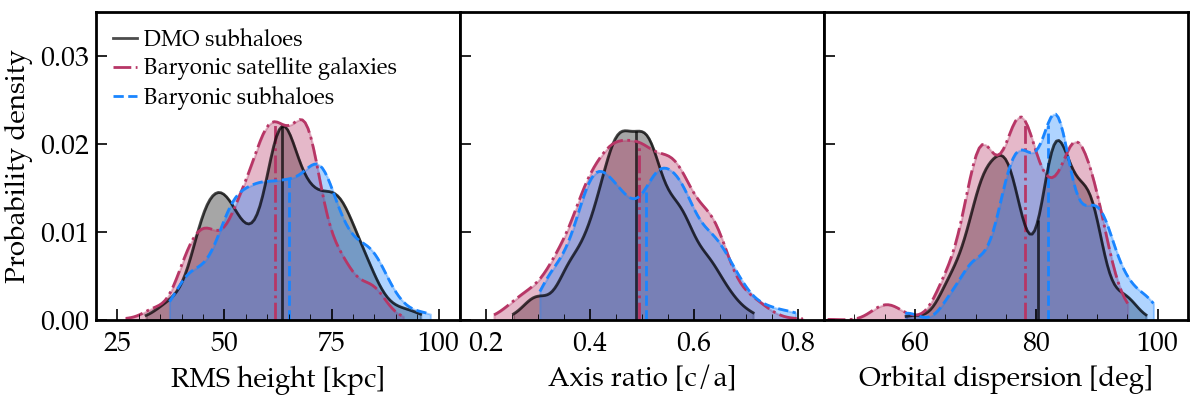}
    \vspace{-6 mm}
    \caption{
    Planarity in baryonic versus dark matter-only (DMO) simulations. 
    We compare the 14 most massive baryonic satellites and subhaloes to DMO subhaloes within $\dlim$.
    We rank order subhaloes by $\Mpeak$ and satellites by $\Mstar$.
    We generate KDEs using 114 snapshots per host over $\zzerotwo$ for each of the hosts and the vertical colored lines are the medians of each distribution.
    All three samples show similar planarity, but a few systems have baryonic satellites with greater kinematic coherence.
    We conclude that there are no significant differences in planarity of baryonic versus DMO simulations.
    }
    \label{fig:dmo}
\end{figure*}

\subsubsection{Baryonic versus dark matter-only simulations}\label{dmo_subsection}

We also ran all of our simulations without baryonic physics, except for one of the isolated hosts, m12z.
We compare planarity of these dark matter-only (DMO) simulations to our baryonic simulations in order to investigate potential baryonic effects on satellite planes, given that many previous studies of planes have used DMO simulations.
We ran the DMO simulations with the same number of DM particles and the same gravitational force softening.
We compare planes in our baryonic simulations to planes in their DMO counterparts by selecting luminous satellites and dark matter subhaloes within $\dlim$.
We choose the 14 most massive object from each sample by rank ordering satellite galaxies by $\Mstar$ and subhaloes using $\Mpeak$.

Figure~\ref{fig:dmo} shows the distributions of planarity for satellite galaxies and subhaloes, both selecting the top 14 subhaloes by $\Mpeak$ and the top satellite galaxies by $\Mstar$, which are not identical samples because of scatter in the $\Mstar - \Mpeak$ relation.
The satellite galaxy distributions (red) are identical to those in Figure~\ref{fig:combined_selections_planes}.
While the three distributions in each panel have slightly different shapes, they have almost the same ranges and medians.
Using a rank ordering selection, the planarity of DMO subhaloes is essentially identical to that of baryonic satellites.
We find that this general result is robust with respect to rank ordering subhaloes by different properties such as $\Mhalo$, $\vpeak$, and $\vcirc$.
The one exception is a small population of baryonic satellites that extend to lower orbital dispersion values during the passage of an LMC-like satellite (see Section~\ref{lmc_subsection}).

These results are surprising in light of the differences in the radial distributions of satellites and subhaloes in our simulations, wherein DMO subhaloes are more radially concentrated around their host than luminous satellites \citep{Samuel2020}.
One might expect to find thinner planes in DMO simulations because subhaloes reside spatially closer to the host halo.
Alternatively, one might expect baryonic simulations to show greater planarity, given that the surviving population is biased to more tangential orbits \citep[e.g.][]{GK2017b}.
\citet{Ahmed2017} observed both a difference in the significance and satellite membership of their planes in baryonic versus DMO simulations of the same four systems.
Satellite membership here refers to whether the satellites contributing to planes belong to the same subhaloes in baryonic and DMO runs of the same systems. 
While we do not explicitly consider differences in satellite membership, we do find that our DMO spatial planes are also typically more significant relative to a statistically isotropic distribution of satellites than their baryonic counterparts, with most having $P<0.5$ during $\zzerotwo$.
However, the significance of DMO kinematic planes is on par with the baryonic simulations.
So while we find that the significance of spatial planes may be slightly overestimated in DMO simulations relative to baryonic simulations, the absolute planarity is not much different from that in baryonic simulations.
If, instead, we select subhaloes at a fixed value of $\mpeaklowlim$, DMO simulations typically have many more subhaloes meeting this criteria.
This difference in number of subhaloes in the plane sample reduces planarity in DMO simulations, because planes with more members are generally less planar \citep{Pawlowski2019}.

\begin{table}
\centering
\caption{
Correlations between planarity in simulations and properties of the host and radial distribution of satellites. 
We select satellite galaxies by rank ordering them by stellar mass at each snapshot and choosing the 14 most massive. 
We measure host halo properties using only dark matter. 
We quote the correlation coefficients ($r$) and $p$-values given by the Spearman correlation test. 
For brevity, we only show correlations with $p<0.1$, though we note that only $p\lesssim0.01$ indicates a significant correlation in our sample.}
\label{tab:corr-table}
\begin{tabular}{llll}
\hline
Planarity metric   & Host/system property                 & $r$ & $p$-value \\
\hline
RMS height         & Host halo concentration              & 0.49 & 0.07    \\
                   & Host $\rm{M}_{*}$       & 0.55 & 0.04    \\
                   & Host halo axis ratio (c/a)           & 0.60 & 0.02    \\
                   & Host $\rm{M}_{*}/\rm{M}_{\rm{halo}}$ & 0.54 & 0.04    \\
                   & $R_{50}$                             & 0.68 & 0.01    \\
\hline
Axis ratio         & Host halo axis ratio (c/a)           & 0.52 & 0.06    \\
\hline
Orbital dispersion & $R_{90}/R_{10}$                      & 0.59 & 0.03    \\
                   & $R_{90}/R_{50}$                      & 0.58 & 0.03   \\
\hline
\end{tabular}
\end{table}

\subsubsection{Correlations between plane metrics and host-satellite system properties}\label{correlations_subsection}

Finally, we explore relationships between satellite planarity and host and satellite system properties, but we find few correlations.
We quantify correlation using the Spearman correlation coefficient ($r$) and $p$-value, applied to the median value of each plane metric and host property over $z=0-0.2$ for all hosts.
None of the correlations that we found are particularly strong, as all have $r<0.7$.
We summarize correlations in Table~\ref{tab:corr-table}, where we only show correlations with $p<0.1$ for brevity.
Only correlations with $p\lesssim0.01$ indicate a statistically significant correlation in our sample, and there is only one correlation meeting this criteria.

We considered four host properties: stellar mass, dark matter halo mass ($\Mtwohm$), stellar-to-total mass ratio, and halo concentration.
Both of the spatial metrics correlate with the host halo axis ratio, such that more triaxial host halos are more likely to have thinner satellite planes.
RMS height is also correlated with host stellar mass, whereby more massive host discs may act to disrupt rather than promote thin planes.
Orbital dispersion does not correlate significantly with any of the host properties.
While there is some evidence for spatial planarity correlating with host halo axis ratio, the correlations are not strong ($0.5<r<0.6$).

We conclude that it is unlikely that host properties drive the formation of satellite planes because we do not find strong and consistent correlations between planarity and host properties.
We also explored the alignment of planes with respect to both the host galaxy disc and the host halo minor axis, but we found no conclusive correlations among our sample.
Given the polar satellite plane around the MW, and the M31 satellite plane being more aligned with the host disc, our results support that we expect no consistent correlation with the disc.
This results agrees with \citet{Pawlowski2019}, who found that satellite plane metrics did not correlate with host properties like halo concentration or halo formation time in dark matter only simulations.

We also test for correlations between planarity and the radial distribution of satellites.
The strongest correlation that we find ($r=0.68$ and $p=0.01$) exists between RMS height and $R_{50}$, the radius enclosing 50 per cent of the satellites.
This correlation may arise from more satellites being near pericenter, rather than actually being flattened into a thin plane, because RMS height is a dimensional quantity (unlike dimensionless axis ratio), so we would expect it to correlate with satellite distances.
We also examine planarity as a function of $\rninetyfifty$ and $\rninetyten$, where $\rninetyfifty$ is the ratio of the distance from the host that encloses 90 per cent of the satellite population to the distance from the host that encloses 50 per cent of the satellite population, and $\rninetyten$ is similarly defined.
These ratios describe the radial concentration of the satellites around their host, and they are the only metrics that significantly correlate with orbital dispersion.
In both cases, more concentrated satellite systems are correlated with less kinematically coherent planes.

We investigated relationships between planarity and properties of major mergers in the histories of the host galaxies.
We adopt the following definition of major merger: a merger occurring during $z=0-3$ with a stellar mass ratio of at least 10 per cent.
Altogether, 10 of the 14 hosts experience at least one major merger.
m12c, m12f, m12r, m12z, Louise, and Remus each have one major merger, while m12m, Thelma, and Romulus each have two, and m12b has a total of three.
Six of the hosts experience mergers that we broadly classify as similar to the Gaia-Enceladus event \citep{Belokurov2018,Helmi2018} by requiring them to have occurred between $\sim8-11$ Gyr ago and to have a stellar mass ratio of $10-30$ per cent.
Two of the hosts (m12m and Thelma) each experience two Gaia-Enceladus type mergers.
Three out of the four hosts with LMC analogues (m12b, m12c, and m12f) experience at least one major merger, and m12b and m12c each experience a merger within the Gaia-Enceladus time window, but with mass ratios (13 and 35 per cent, respectively) just outside of our nominal range.
We tested for correlations between planarity during $\zzerotwo$ and the number of major mergers per host, the timing of the last major merger, and the mass ratio of the last major merger, but found no significant correlations.
The strongest correlations were between orbital pole dispersion and merger properties, but the maximum correlation coefficient was only $\sim0.4$ with a $p$-value of $\sim0.2$, indicating overall weak correlations between planarity and major mergers in the host.
While there is perhaps some evidence for a correlation between LMC analogues (arguably the most important factor in creating satellite planes) and major mergers, we conclude that past major mergers do not have a strong independent influence on planarity.

\section{Summary and Discussion}\label{discussion}

We explored the incidence and origin of planes of satellite galaxies in the FIRE-2 simulations, using satellites around 14 MW/M31-mass galaxies over $\zzerotwo$. 
We compared to and provided context for satellite planes in the Local Group, including all satellites with $\mstarlowlim$ around the MW and within the PAndAS survey of M31.
We summarize our main results as follows.

\subsection{Rareness of Planes}
\begin{itemize}
    \item MW-like planes exist in our simulations, but they are relatively rare among our randomly selected $\sim10^{12}\,M_{\odot}$ halos at $\zzerotwo$: planes at least as thin or coherent as the MW's plane occur in $\sim1-5$ per cent of all snapshots, and planes as thin and coherent according to \textit{spatial and kinematic metrics simultaneously} occur in $\sim 0.3$ per cent of snapshots.
    \item \textit{However}, if we select halos that feature a LMC-mass satellite analogue near its first pericentric passage, then the frequency of MW-like or thinner planes dramatically increases to $7-16$ per cent, with $\sim 5\%$ at least as thin as the MW plane by spatial and kinematic metrics simultaneously.
    \item If we consider M31's satellite population as a whole, the planarity of satellites around M31 is common in our simulations. By every spatial or kinematic (or simultaneous) measure we consider, M31's satellites lie within $\sim1\sigma$ of the median of randomly selected halos of similar mass that we simulated.
    \item Most of our simulations are not significantly planar relative to a statistically isotropic distribution of satellites.
\end{itemize}

\subsection{Physical Origins of Planes}
\begin{itemize}
    \item Most MW-like thin satellite planes are transient and last $<500$\,Myr in our simulations. However, the presence of an LMC satellite analogue near pericenter produces longer lifetimes of $\sim 0.7-3\,$Gyr. More generically flattened satellite systems survive for up to $\sim2-3$\,Gyr, even without requiring a massive satellite like an LMC analogue.
    \item We do not find significant differences in planarity of satellites around hosts in Local Group-like pairs versus isolated hosts.
    \item Dark matter-only (DMO) simulations show no significant differences in planarity compared to their baryonic-simulation counterparts, when selecting a fixed number of satellites in each sample.
    \item Correlations between plane thickness and other satellite population properties (radial concentration) or host properties (mass, concentration, size, axis ratio) are generally modest or weak. Plane thickness is generally larger for more radially extended satellite distributions, as expected. The one property that strongly correlates with the presence of spatially thin and kinematically coherent planes is the presence of an LMC analogue near first pericentric passage.
\end{itemize}

\subsection{Observational and Selection Effects}
\begin{itemize}
    \item Plane metrics can be sensitive to the satellite selection method in simulations and observations. Selecting just the 14 satellite galaxies with the highest stellar mass in the simulation produces thinner planes compared to selecting \textit{all} satellites with $M_{\ast}>10^{5}\,M_{\odot}$, because the latter tends to select more satellites, which produces thicker planes. 
    \item Incompleteness from the inability to see through the host galaxy disc (as in the MW) can increase the probability of measuring MW-like spatial planes by as much as a factor $\sim 10$. This bias is opposite in sign but much smaller for kinematic planes.
    \item We have \textit{not} corrected in any of our analysis for any `look-elsewhere' effects, including the choice to look for `planes', the choice of definition of `plane', sample selection, number of satellites, etc. These corrections only would \textit{decrease} the statistical significance of the observed planes, as outliers from simulations.
\end{itemize}

\subsection{Discussion}

Though only $1-2$ per cent of snapshots for all 14 hosts during $\zzerotwo$ contain satellite planes at least as thin as the MW's, we do not interpret this as a strong tension with $\lcdm$ cosmology.
Instead, we identify the mere presence of MW-like planes in the simulations as evidence that cosmological simulation indeed can form thin planes of satellites, as long as they have adequate mass and spatial resolution.
We find that planes are much more common in the presence of LMC analogues, as suggested by \citet{Li2008} and \citet{DOnghia2008}, which provides evidence that future work should prioritize comparing the MW against simulations with an LMC analogue.
Considering the entire M31 satellite population, M31-like satellite planes are common in our simulations, and combined with the fact that our simulations are only marginally more planar than a statistically isotropic distribution of satellites, this may indicate that M31's satellites as a whole are not significantly planar.
Our most promising result points to the presence of the LMC near first pericenter as a likely primary driver of planarity.
The lack of strong correlations between planarity and other properties of the host-satellite systems leaves us with few other physical explanations for the MW's highly coherent satellite plane.
If our simulations are representative of the MW, then the observed MW plane is likely to be a temporary effect that will wash out in subsequent orbits of the LMC \citep{Deason2015}.

\citet{Pawlowski2020} report similar percentages ($\sim2-3$ per cent) of hosts with MW-like planes in the IllustrisTNG simulations, but contrary to our own conclusion, they claim that this does in fact constitute a challenge to $\lcdm$.
Comparing the percentages in \citet{Pawlowski2020} to those that we obtain in this study is a readily understandable synthesis of the two studies, but we now highlight a few key differences in the underlying data sets (both observed and simulated) and analysis that warrant a more nuanced comparison of our work with that of \citet{Pawlowski2020}. 
Our observational sample includes three additional MW satellites (Crater II, Antlia II, and Canes Venatici I) that meet our nominal stellar mass criteria ($\mstarlowlim$).
\citet{Pawlowski2020} may have excluded these satellites because they are borderline cases of `classical' dwarfs given their diffuse morphology and/or stellar masses, or perhaps because some were discovered only recently and thus not ideal for the historical comparison in that work.

In calculating orbital pole dispersion in simulated systems, \citet{Pawlowski2020} sample from each host's satellites to select the most aligned subsample, something that we do not explore here.
The use of a `most-aligned' sub-sample could lead us to measure smaller orbital pole dispersions in our systems (and hence a higher fraction of snapshots with MW-like kinematic planes), but we are limited in sample size given our stellar mass criteria.
The percentages from \citet{Pawlowski2020} represent robust statistical significances, because their data are comprised of over 1000 independent host systems, whereas our data include multiple snapshots for each of only 14 independent host systems.
In comparison, our measured percentages are not strict statistical significances.

Furthermore, our claim that Local Group satellite planes are not a strong challenge to $\lcdm$ rests on two main conclusions from our work that are not in \citet{Pawlowski2020}: the presence of an LMC analogue makes measuring a MW-like plane more likely (and indicates a system that is a better match to the MW), as well as the commonality of M31-like planes.
We conclude that, because MW-like planes are less rare when we match the MW's satellite population more precisely than in previous studies of satellite planes, and because we only compare to one observed system in this case, that observed satellite planes do not constitute a strong challenge to $\lcdm$.
The rareness of satellite planes remains an interesting topic, but we maintain that a \textit{strong} challenge to $\lcdm$ cosmology requires \textit{strong} evidence of rarity (as opposed to mere uncertainty), which we do not find in this work.

We have deliberately approached our analysis of satellite planes as agnostically as we can.
In choosing a fixed number of satellites for our nominal selection method, we have tried to both show the clearest comparisons between our simulations and LG observations, as well as mitigate the confounding effects of correlations between $\nsat$ and planarity.
Further studies of the most-planar subsamples of simulated satellites, as examined in \citet{Pawlowski2013} and extended in \citet{SantosSantos2020}, may yield more insight into the nature of satellite planes.
We defer an analysis of satellite sub-samples to future work.

We also have not yet considered a comparison to satellite systems outside of the LG.
There is evidence for satellite planes outside of the LG around Centaurus A \citep{Muller2018}, and recent studies have examined planarity around hosts in SDSS \citep{Ibata2014b,Brainerd2020} and the SAGA survey \citep{Mao2020}.
Connecting LG hosts to a statistical sample of similar hosts will be crucial in evaluating the significance of planar alignments and the validity of proposed formation mechanisms, demonstrating the need for large surveys with e.g., the Nancy Grace Roman Space Telescope, which promises to significantly augment the observational sample of MW analogues.
LG galaxies are also aligned with large scale structure, along a local sheet, which is not captured in our simulations and may play a part in the formation of satellite planes \citep{Neuzil2020}.
Simulations that can accurately reproduce this large scale structure may offer new insight into satellite planes \citep{Libeskind2020}.

\section*{Acknowledgements}

We thank Marcel Pawlowski for insightful comments and discussion that improved this manuscript.

This research made use of Astropy,\footnote{http://www.astropy.org} a community-developed core Python package for Astronomy \citep{astropy:2013, astropy:2018}, the IPython package \citep{ipython}, NumPy \citep{numpy}, SciPy \citep{scipy}, Numba \citep{numba}, and matplotlib, a Python library for publication quality graphics \citep{matplotlib}.

JS, AW, and SC received support from NASA through ATP grants 80NSSC18K1097 and 80NSSC20K0513; HST grants GO-14734, AR-15057, AR-15809, and GO-15902 from the Space Telescope Science Institute (STScI), which is operated by the Association of Universities for Research in Astronomy, Inc., for NASA, under contract NAS5-26555; a Scialog Award from the Heising-Simons Foundation; and a Hellman Fellowship.
We performed this work in part at the Aspen Center for Physics, supported by NSF grant PHY-1607611, and at the KITP, supported NSF grant PHY-1748958.
PFH was provided by an Alfred P. Sloan Research Fellowship, NSF grant \#1715847 and CAREER grant \#1455342, and NASA grants NNX15AT06G, JPL 1589742, 17-ATP17-0214.
MBK acknowledges support from NSF CAREER award AST-1752913, NSF grant AST-1910346, NASA grant NNX17AG29G, and HST-AR-15006, HST-AR-15809, HST-GO-15658, HST-GO-15901, and HST-GO-15902 from STScI.
CAFG was supported by NSF through grants AST-1517491, AST-1715216, and CAREER award AST-1652522, by NASA through grant 17-ATP17-0067, and by a Cottrell Scholar Award from the Research Corporation for Science Advancement.
We ran simulations using the Extreme Science and Engineering Discovery Environment (XSEDE) supported by NSF grant ACI-1548562, Blue Waters via allocation PRAC NSF.1713353 supported by the NSF, and NASA's HEC Program through the NAS Division at Ames Research Center.

\section*{Data Availability}

Full simulation snapshots at $z = 0$ are available for m12i, m12f, and m12m at \url{ananke.hub.yt}.
The publicly available software packages used to analyze these data are availble at: \url{https://bitbucket.org/awetzel/gizmo\_analysis}, \url{https://bitbucket.org/awetzel/halo\_analysis}, and \url{https://bitbucket.org/awetzel/utilities}.



\bibliographystyle{mnras}
\bibliography{satellite}


\appendix

\section{Alternative proper-motion measurements}
\label{proper_motion_appendix}

\citet{Pawlowski2020} compiled a `best-available' sample of proper motions for their analysis of satellite planes by choosing measured proper motions of the 11 most massive classical dwarf MW satellites from the literature that have the smallest uncertainties.
Compared to our sample, \citet{Pawlowski2020} do not include Crater II, Antlia II, or Canes Venatici I in their sample, likely for consistency with past analyses and because the stellar masses of these satellites are close to the lower limit for classical dwarfs ($\sim10^5 \Msun$) and some of them show evidence for tidal disruption.
Of the satellites that both of our samples have in common, all of the proper motions are approximately the same except for that of Leo II: \citet{Pawlowski2020} uses a Leo II proper motion from \citet{Piatek2016} based on HST data, which is significantly different from the value we use in magnitude, direction, and uncertainty.
Any other differences in proper motions between the two samples are not significant enough to alter our analysis.

\begin{figure*}
    \begin{multicols}{2}
	\includegraphics[width=0.45\textwidth]{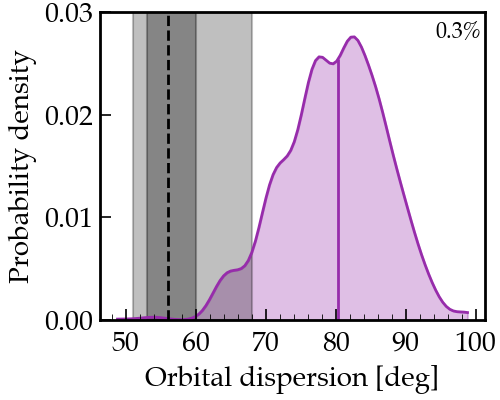}\par
	\includegraphics[width=0.45\textwidth]{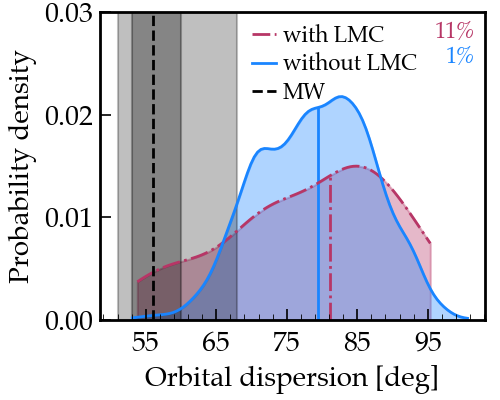}\par
	\end{multicols}
	\vspace{-6 mm}
    \caption{Effects of an alternative proper motion sample for the observed MW satellites. Note that the redshift ranges for the left and right plots are not the same ($\zzerotwo$ and $z\sim0-0.7$, respectively).
    \textit{Left:} Same as the right panel of Figure~\ref{fig:mw_completeness_planes}, but with the `best-available' proper motion for Leo II from HST observations \citep{Piatek2016}.
    Using the best-available proper motion for Leo II shifts the median orbital pole dispersion for the MW from $60\deg$ to $56\deg$, and also decreases the scatter to larger angles.
    This has the effect of reducing the fraction of snapshots with MW-like (at or below the MW's upper 68 per cent limit) planes from 5 per cent to 0.3 per cent, however, we note that almost 6 per cent of snapshots lie at or below the MW's upper 95 per cent limit.
    \textit{Right:} Same as the right panel of Figure~\ref{fig:lmc_passages}, but using the proper motions described above.
    While there are fewer snapshots that are MW-like overall, the enhancement in the fraction of snapshots that are MW-like during LMC-like pericenter passages relative to the general sample of snapshots is still evident.}
    \label{fig:proper_motion_appendix}
\end{figure*}

Using the \citet{Piatek2016} proper motion and sampling from the given uncertainties, we measure a narrower range and lower median orbital pole dispersion ($56\deg$ versus $60\deg$) for MW satellites.
Figure~\ref{fig:proper_motion_appendix} shows the main effect on our analysis: fewer simulation snapshots have MW-like orbital pole dispersions.
Only 0.3 per cent of snapshots during $\zzerotwo$ are at least as kinematically coherent as the MW satellite plane, whereas this value was previously 5 per cent using the Gaia proper motion for Leo II. 
Most importantly, the main conclusion of Section~\ref{lmc_subsection} still holds: MW-like planes are still more likely during the first pericentric passage of an LMC-like satellite, and using the best-available proper motion sample actually enhances that result.
Using the best-available proper motion sample, MW-like kinematic planes are 11 times more likely in the presence of our LMC analogues, versus $\sim3$ times more likely with our original proper motion sample.
While the best-available proper motion sample makes MW-like kinematic planes more rare in our simulations, it does not affect our results on spatial thinness of planes and it does not qualitatively change any of our other results.
A definitive observational proper motion for Leo II would enable us to perform an even more robust analysis of the MW's satellite plane in comparison to cosmological simulations.

\section{Comparison to statistically isotropic realizations}
\label{isotropic_appendix}

\begin{figure*}
    \includegraphics[width=\textwidth]{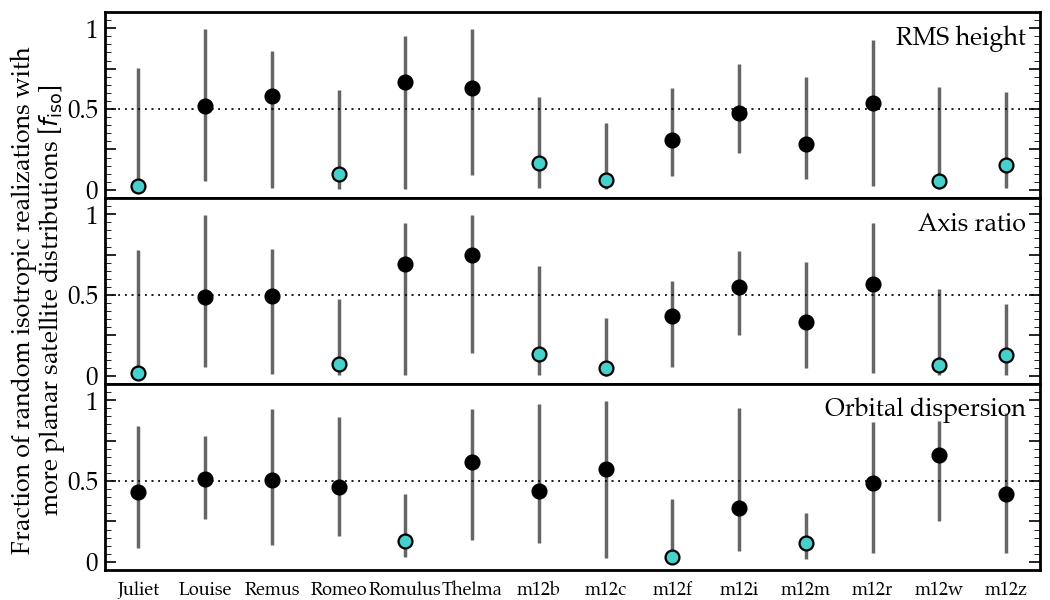}
    \vspace{-2 mm}
    \caption{
    The significance of simulated and observed plane metrics from Figure~\ref{fig:mw_completeness_planes} relative to $\nitertenk$ statistically isotropic realizations of satellite positions (keeping radial distance fixed) and velocities (considering only their directions). 
    For each simulated host we plot the median and 95 per cent scatter during $\zzerotwo$ in the fraction ($f_{\rm iso}$) of isotropic realizations that are more planar than the true plane.
    We consider hosts with median $f_{\rm iso}\leq0.25$ and lower 95 per cent limit $f_{\rm iso}\leq0.05$ to have significant planes (blue).
    The MW's plane is highly significant relative to its statistically isotropic distribution, both spatially ($f_{\rm iso}=0.003$) and kinematically ($f_{\rm iso}=0.005$).
    About half of the simulated hosts (Juliet, Romeo, m12b, m12c, m12w, and m12z) have significant spatial planes, and only three (Romulus, m12f, and m12m) have significant kinematic planes during $\zzerotwo$.
    None of the simulated hosts are significant in both a spatial and kinematic sense, and most hosts are consistent with a statistically isotropic distribution.
    While MW-like planes do occur in the simulations, they are not as significant as the MW's plane.
    }
    \label{fig:isotropic_appendix}
\end{figure*}

In Figure~\ref{fig:isotropic_appendix}, we provide a visual representation of how planar each host's satellite system is relative to $\nitertenk$ statistically isotropic random realizations of satellite positions and velocities.
We describe this calculation in detail in Sections~\ref{isotropic_subsection} and~\ref{statistical_subsection}.

\bsp	
\label{lastpage}
\end{document}